\documentclass[12pt]{article}

	\usepackage{geometry}
	\usepackage[T1]{fontenc}
	\usepackage{times}
	\usepackage{mathptmx}
	\usepackage{titlesec}
	\titleformat{\section}{\raggedright\normalsize\bfseries}{\thesection.}{0.75em}{}
	\titleformat{\subsection}{\raggedright\normalsize\itshape}{\thesubsection.}{0.75em}{}
	\usepackage{caption2}

	\usepackage{amsfonts,amssymb}
	\usepackage{indentfirst}
	\usepackage{latexsym}
	\usepackage{stmaryrd}
	\usepackage{amsmath,amsbsy,mathtools}
	\usepackage{graphicx}
	\allowdisplaybreaks[4]
	\usepackage{hyperref}
	\usepackage[authoryear,colon]{natbib}
	\bibpunct{(}{)}{;}{a}{}{,}
	\usepackage{booktabs}
	\usepackage[ruled]{algorithm2e}
	
	\usepackage{authblk}

	\let\origintitle\title
	\def\title#1{\origintitle{\textbf{#1}}}
	\newenvironment{keywords}{\par\noindent\footnotesize\textbf{Keywords: }}{\par}
	\newenvironment{MSC}{\par\noindent\footnotesize\textbf{MSC2010: }}{\par}

	\usepackage{color}

	\newcommand{\comment}[1]{\relax}
	
	\usepackage{enumitem}
	\usepackage{float}
	\usepackage{subfigure}
	
	\newcommand{\reftab}[1]{Table~\ref{#1}}
	\newcommand{\reffig}[1]{Figure~\ref{#1}}
	\newcommand{\refsect}[1]{Section~\ref{#1}}
	\newcommand{\refeqt}[1]{\eqref{#1}}
	\newcommand{\refeqttageqts}{}
	
	\newcommand{\br}[1]{(#1)}
	\newcommand{\bbr}[1]{\left(#1\right)}
	\def\bracketclassical{1}
	\ifnum \bracketclassical=0
		\newcommand{\brlv}[2]{\br{#2}}
		\newcommand{\bbrlv}[2]{\bbr{#2}}
	\fi
	\ifnum \bracketclassical=1
		\newcommand{\brlv}[2]{\ifnum #1<2 (#2) \else \ifnum #1=2 [#2] \else \{#2\} \fi \fi}
		\newcommand{\bbrlv}[2]{
			\ifnum #1<2 \left(#2\right) \else \ifnum #1=2 \left[#2\right] \else \left\{#2\right\} \fi \fi
			}
	\fi

	\def\squiggle#1{\mathcal{#1}}
	\newcommand{\omd}{\cdots}

	\newcommand{\setdef}[2]{\left\{ #1 : #2 \right\}}

	\newcommand{\setNp}{\mathbb{N}^{+}}
	\newcommand{\setR}{\mathbb{R}}

	\newcommand{\MatI}{\mathbf{I}}

	\newcommand{\diff}{\mathrm{\,d}}

	\newcommand{\Exp}{\mathrm{E}}
	\newcommand{\Var}{\mathrm{Var}}
	\newcommand{\DistrU}{\mathrm{U}}
	\newcommand{\DistrN}{\mathrm{N}}
	\newcommand{\DistrBi}{\mathrm{Bi}}

	\newcommand{\IID}{\text{i.i.d.}}
	\newcommand{\infrac}[2]{{#1}/{#2}}
	\newcommand{\outfrac}[2]{\dfrac{#1}{#2}}
	\newcommand{\insumop}[3]{\sum\nolimits_{#1}^{#2}{#3}}
	\newcommand{\outsumop}[3]{\sum\limits_{#1}^{#2}{#3}}

	\newcounter{LZflowdiagramstep}
	\newenvironment{LZflowdiagram}{
		\setcounter{LZflowdiagramstep}{0}
		\newcommand{\flowstep}[1]{
			\ifnum\value{LZflowdiagramstep}>0
				\par\noindent $\downarrow$
			\fi
			\par\noindent
			\fbox{\parbox[t]{0.75\columnwidth}{##1}}
			\addtocounter{LZflowdiagramstep}{1}
		}
	}{
		\par
	}
	
	\newcounter{refsubfigure}
	\newcommand{\subfiglabel}[1]{
		\renewcommand{\therefsubfigure}{\thefigure\thesubfigure}
		\refstepcounter{refsubfigure}
		\label{#1}
	}

	\newenvironment{stepparenumerate}
		{
			\begin{enumerate}[fullwidth,itemindent=\parindent,listparindent=\parindent,itemsep=0ex,partopsep=0ex,parsep=0ex,topsep=0ex]

		}
		{\end{enumerate}}

	\def\UKUSswitch{UK}
	\newcommand{\UKUS}[1]{%
		\csname UKUSspell_#1_\UKUSswitch \endcsname%
		\expandafter \ifx \csname UKUSspell_#1_\UKUSswitch \endcsname \relax ??\fi%
	}
	\newcommand{\UKUSadd}[3]{
		\def\tempUKUSswitch{UK}
		\expandafter\gdef\csname UKUSspell_#1_\tempUKUSswitch \endcsname{#2}
		\def\tempUKUSswitch{US}
		\expandafter\gdef\csname UKUSspell_#1_\tempUKUSswitch \endcsname{#3}
	}
	\UKUSadd{is}{is}{iz}
	\UKUSadd{our}{our}{or}
	\UKUSadd{re}{re}{er}
	\UKUSadd{red}{red}{ered}
	
	\def\citeprefixswitch{authoryear} 
	\newcommand{\citeprefix}[3]{%
		\def\tempciteprefixswitch{authoryear}%
		\ifx \citeprefixswitch \tempciteprefixswitch #1 \fi%
		\def\tempciteprefixswitch{numbers}%
		\ifx \citeprefixswitch \tempciteprefixswitch #2 \fi%
		\def\tempciteprefixswitch{super}%
		\ifx \citeprefixswitch \tempciteprefixswitch #3 \fi%
	}

	\newcommand{\inmathbreaklinespace}{\linebreak[4]}
	\newcommand{\nns}[1]{\squiggle{Z}_{#1}}
	\newcommand{\vecs}[1]{\boldsymbol{#1}}
	
	\newcommand{\round}{\mathrm{round}}
	\newcommand{\critSSD}{\mathrm{SSD}}
	\newcommand{\measure}{\mathbf{m}}

	\newcommand{\region}{\squiggle{R}}
	\newcommand{\rvpos}{\xi}
	\newcommand{\rvshift}{\delta}
	\newcommand{\nvpos}{\vecs{x}}
	\newcommand{\desUD}{D}
	\newcommand{\entUD}{\vecs{q}}
	\newcommand{\sspop}{\mathrm{P}}
	\newcommand{\ssinf}{\mathrm{I}}
	\newcommand{\ssdiag}{\mathrm{D}}
	\newcommand{\total}{N}
	\newcommand{\totalpop}{\total_{\sspop}}
	\newcommand{\totalinf}{\total_{\ssinf}}
	\newcommand{\totaldiag}{\total_{\ssdiag}}
	\newcommand{\denst}{f}
	\newcommand{\denstpop}{\denst_{\sspop}}
	\newcommand{\denstinf}{\denst_{\ssinf}}
	\newcommand{\denstdiag}{\denst_{\ssdiag}}
	\newcommand{\denstvarker}[1]{\tilde{\denst}#1}
	\newcommand{\denstsam}{\varphi}
	\newcommand{\weightsamsam}{w}
	\newcommand{\sizesam}{n}
	\newcommand{\sizepos}{r}
	\newcommand{\sizeUD}{M}
	\newcommand{\alloc}{\nu}
	\newcommand{\cost}{c}
	\newcommand{\coefcost}{\gamma_{\cost}}
	\newcommand{\rate}{\rho}
	\newcommand{\rateinf}{\rate_{\ssinf}}
	\newcommand{\ratesam}{\rate_{\mathrm{S}}}
	
	\newcommand{\coefdenstinf}{\gamma}
	\newcommand{\coefsizesampos}{\eta}
	\newcommand{\testres}{\tau}

	\newcommand{\rvshiftbat}[1]{\rvshift^{(#1)}}
	\newcommand{\desUDbat}[1]{\desUD^{(#1)}}
	
	\newcommand{\weightsamsambat}[1]{\weightsamsam^{(#1)}}
	\newcommand{\distrsamsambat}[1]{\squiggle{P}^{(#1)}}

	\newcommand{\denstpoppos}[1]{\denst_{\sspop,#1}}
	\newcommand{\denstinfpos}[1]{\denst_{\ssinf,#1}}
	\newcommand{\denstdiagpos}[1]{\denst_{\ssdiag,#1}}
	\newcommand{\sizesampos}[1]{\sizesam_{#1}}
	\newcommand{\coefdenstinfpos}[1]{\gamma_{#1}}

	\newcommand{\estdenstinf}{\hat{\denst}_{\ssinf}}
	\newcommand{\esttotalinf}{\hat{\total}_{\ssinf}}
	\newcommand{\estcoefdenstinf}{\check{\gamma}}
	\newcommand{\estdenstinfrough}{\check{\denst}_{\ssinf}}
	\newcommand{\estdenstvarkerrough}[1]{\check{\tilde{\denst}}#1}
	
	\newcommand{\USAdatediag}[1]{December 27th, 2020}
	\newcommand{\USAdateinf}[1]{April 22nd, 2021}
	\newcommand{\USAdatediags}[1]{2020-12-27}
	\newcommand{\USAdateinfs}[1]{2021-04-22}
	
	\newcounter{refSM}
\begin{document}

	\title{A sampling scheme for estimating the prevalence of a pandemic}
	\author[1]{Ze Liu}
	\author[1]{Siyu Yi}
	\author[2]{Jianghu (James) Dong}
	\author[1]{Min-Qian Liu}
	\author[1]{Yongdao Zhou \thanks{ydzhou@nankai.edu.cn}}
	\affil[1]{School of Statistics and Data science, LPMC \& KLMDASR,
		NanKai University, Tianjin, China}
	\affil[2]{Department of Biostatistics, College of Public Health,
		University of Nebraska Medical Center, Omaha, Nebraska, U.S.A.}
	\date{}
	\maketitle

	\begin{abstract}
The spread of COVID-19 makes it essential to investigate its prevalence. In such investigation research, as far as we know, the widely-used sampling methods 		didn't use the information sufficiently about the numbers of the previously diagnosed cases, which provides a priori information about the true numbers of infections. This motivates us to develop a new, two-stage sampling method in this paper, which utilises the information about the distributions of both population and diagnosed cases, to investigate the prevalence more efficiently. The global likelihood sampling, a robust and efficient sampler to draw samples from any probability density function, is used in our sampling strategy, and thus, 		our new method can automatically adapt to the complicated distributions of population and cases. Moreover, the corresponding estimating method is simple, 		which facilitates the practical implementation. Some recommendations for practical implementation are given. Finally, several simulations and a practical example verified its efficiency.
	\end{abstract}
	
	
	\begin{MSC}
		62D05, 62P10, 65C05.
	\end{MSC}

	\section{Introduction}\label{sect_intro}
	

			COVID-19 broke out at the end of 2019
			and was declared a global pandemic by
\cite{2020_WHO_0}.
			All countries around the world were severely affected,
			especially the United States of America (USA),
			where the numbers of confirmed cases and deaths increased rapidly
			from March 2020 to April 2021 \citep{2020_CDC_web}.
			It is essential for the government and health institutions to monitor COVID-19
			and control the pandemic by making practical and reasonable plans.
			
			Since some people infected by SARS-CoV-2 are asymptomatic \citep{bib_pp_39},
			one main difficulty with the pandemic is that
			the cumulative number of diagnosed cases cannot represent the number of infections.
			Many countries have made efforts to investigate the prevalence of COVID-19
			\citep{bib_pp_32, bib_pp_35, bib_pp_36, bib_pp_37, bib_pp_38, bib_pp_41,
			2020_Sood_JAMA, bib_pp_15};
			however,
			many of these investigations
			\citep{bib_pp_35, bib_pp_37, bib_pp_38, 2020_Sood_JAMA}
			only focused on one or several hotspot(s) instead of the whole country.
			\comment{while \cite{bib_pp_15, bib_pp_36, bib_pp_32, bib_pp_24}
			estimated the nationwide prevalences.}
			In addition,
			when investigating the prevalence nationwide,
			restricted by the costs,
			only a small part of the whole population can be investigated,
			especially when the country has a broad territory area.
			Therefore, it is important to develop some appropriate sampling strategies.
			The use of convenience samples
			\citep{bib_pp_35, bib_pp_37, bib_pp_38, bib_pp_41,
			bib_pp_21}
			is not proper because they are prone to the selection bias,
			and thus, problematic.
			Some other literature work
			\citep{bib_pp_14, bib_pp_18, bib_pp_25, bib_pp_26, bib_pp_28}
			drew samples from some representative databases
			--- for example, the medical insurance databases.
			This strategy can reduce the cost of the survey,
			but the representativeness of such samples
			depends on the representativeness of the database.
			Another popular sampling strategy used in the investigations of prevalence is
			the multi-stage stratified sampling, or some variants of it.
			Some recent studies
			\citep{bib_pp_30, bib_pp_36, bib_pp_15,
			bib_pp_16, bib_pp_19, bib_pp_20, bib_pp_24, bib_pp_27}
			have used these kinds of sampling methods.
			Compared with the simple random sampling,
			the variance of the estimator obtained by the stratified sampling is usually smaller.
			However, this method depends heavily on the construction of the strata
			in which the inter-homogeneity is required,
			and sometimes a well-designed stratified sampling strategy
			may lead to a very complex analysis procedure.
			
			Intuitively,
			the distribution of the number of cumulative cases
			provides priori information about the true situation of infections,
			and, therefore, can help with the sampling survey to improve the efficiency.
			However, none of the above research util\UKUS{is}ed this priori information sufficiently.
			In this paper,
			we propose a new sampling strategy for estimating the total number of infections nationwide.
			The main feature of this method is that
			it can flexibly and efficiently util\UKUS{is}e the information about
			the distributions of both population and diagnosed cases.
			Compared with the stratified multi-stage sampling,
			our method is more flexible to adapt to
			various complicated distributions of population and cases.
			The implementation and the corresponding estimating methods
			of this sampling strategy are also easier.
			There are two stages in the sampling strategy:
			first, determining the sampling positions according to some probability density,
			and then sampling from these positions.
			The main focus of this paper is on the first stage,
			in which
			the probability density may be multimodal and complicated.
			In this situation,
			some well-known methods to sample from a general probability density function,
			including the Markov Chain Monte Carlo (MCMC) method,
			e.g. Metropolis-Hastings (MH) algorithm \citep{1970_Hastings_Biometrika},
			and the sampling/importance resampling (SIR) method \citep{1987_Rubin_JASA},
			as well as its variants \citep{bib_pp_46, bib_pp_45},
			may have a bad performance.
			The reason is that those methods can't adapt to various kinds of sampling densities
			(the MCMC method is easy to get stuck at some peak of the density function
			when the sampling density is multimodal,
			and the performance of the SIR depends heavily on the choice of the proposal distribution
			and the quality of the initial samples from the proposal distribution).
			To overcome these problems,
			\cite{2015_Wang_conf} proposed a new method,
			called the global likelihood sampling (GLS),
			which is also used in the first stage of our proposed sampling strategy.
			As for the second stage,
			any advanced sampling strategy can be used,
			but for simplicity,
			we only consider the simple random sampling method in this paper.
			
		
			The rest of this paper is organ\UKUS{is}ed as follows.
			In \refsect{sect_prelim},
			we describe the problem and propose the basic approach.
			The GLS algorithm is also described in this section.
			In \refsect{sect_optimal},
			the optimal settings of our sampling strategy are derived.
			The complete sampling strategy and the corresponding estimating method
			are given in \refsect{sect_prac},
			as well as some suggestions for practical implementation.
			Some numerical simulations are conducted in \refsect{sect_simu},
			in order to find the robust setting of a parameter in our method,
			and to show the efficiency of our method.
			To further explain our method,
			a practical example is presented in \refsect{sect_USA}.
			Finally, \refsect{sect_conclusions} concludes this paper.
			Some additional remarks, details and simulations
			are provided in the supplementary material.
		
	\section{Preliminaries}\label{sect_prelim}
		
		\subsection{Basic Approach}\label{subsect_prelim_basic}
			
			Suppose a pandemic is spreading over a region $\region \subseteq \setR^{2}$,
			and we want to know its total prevalence over the region $\region$.
			The total population of this region,
			the cumulative number of diagnosed cases,
			and the cumulative number of real infections
			are denoted by $\totalpop$, $\totaldiag$ and $\totalinf$, respectively.
			The corresponding densities of population, cases and infections in $\region$ are
			$\denstpop$, $\denstdiag$ and $\denstinf$, respectively,
			and
			\begin{align*}
				\int\nolimits_{\region} \denst_{i}(\nvpos) \diff \nvpos &= \total_{i}
			\end{align*}
			with $i$ being $\sspop$, $\ssdiag$ and $\ssinf$.
			The information about the population, $\totalpop$ and $\denstpop$,
			and that about the diagnosed cases, $\totaldiag$ and $\denstdiag$,
			are usually known,
			but the information about the real infections, $\totalinf$ and $\denstinf$,
			is hard to get,
			which is just what we are interested in.
			Since the total prevalence is $\infrac{\totalinf}{\totalpop}$,
			estimating the total prevalence is equivalent to
			estimating the total number of infections $\totalinf$.
			Therefore, for convenience,
			our goal is to estimate $\totalinf$ by a sampling survey.
			
			Denote $\nns{k} := \{1, \omd, k\}$ for any $k \in \setNp$
			where $\setNp$ is the set of all positive integers.
			Since $\totalinf$ is the integral of $\denstinf$ on $\region$,
			a popular way to approximate $\totalinf$ is the Monte Carlo method:
			let $\denstsam$ be a probability density function on $\region$
			and $\rvpos_{1}, \omd, \rvpos_{\sizepos}$
			be $\sizepos$ independent and identically distributed ($\IID$) samples from $\denstsam$,
			then the sample mean
			$\sizepos^{-1} \insumop{i=1}{\sizepos}
			{\infrac{\denstinf(\rvpos_{i})}{\denstsam(\rvpos_{i})}}$
			is an unbiased estimator of $\totalinf$.
			Hereafter,
			we call $\denstsam$ the sampling density
			and call $\rvpos_{1},\omd,\rvpos_{\sizepos}$ the sampling positions.
			Since $\denstinf$ is unknown,
			we have to estimate the values of $\denstinf$ at $\rvpos_{1}, \omd, \rvpos_{\sizepos}$.
			Therefore, our sampling survey consists of two stages:
			first,
			determining the sampling positions $\rvpos_{1},\omd,\rvpos_{\sizepos}$;
			second,
			selecting the samples,
			i.e., the people who will receive the tests, at each sampling position
			to estimate the values of $\denstinf$ there.
			The main focus of this paper is on the first stage,
			and the simple random sampling is adopted in the second stage.
			Other advanced sampling methods can also be used in the second stage,
			which is problem-dependent.
			
			\comment{First consider the second stage
			in which our goal is to estimate the values of $\denstinf$ at several sampling positions.}
			Denote the local prevalence function over $\region$ by
			$\rateinf := \infrac{\denstinf}{\denstpop}$.
			Let $\nvpos$ be an arbitrary fixed position in the region $\region$,
			and $\sizesampos{\nvpos}$ be the fixed sample size at $\nvpos$.
			Then, $\sizesampos{\nvpos}$ people from the sampling position $\nvpos$ are tested.
			For each $i \in \nns{\sizesampos{\nvpos}}$,
			let $\testres_{i} = 1$ if the $i$-th person was infected,
			and $\testres_{i} = 0$ otherwise.
			Intuitively,
			one person's possibility of infection is associated with
			the infection status of others in that person's household/neighb\UKUS{our}hood/community.
			This correlation can be character\UKUS{is}ed by the local prevalence function $\rateinf$.
			Moreover, $\sizesampos{\nvpos}$ is usually very small
			compared to the population around the position $\nvpos$.
			Hence, we can consider, approximately, that
			$\testres_{1}, \omd, \testres_{\sizesampos{\nvpos}}
			\stackrel{\IID}{\sim} \DistrBi(1, \rateinf(\nvpos))$,
			where $\DistrBi$ represents the binomial distribution.
			An unbiased estimator of $\denstinf(\nvpos)$ is
			\begin{align}\label{eq_estdenstinf}
				\estdenstinf(\nvpos)
				:= \denstpop(\nvpos) \cdot \outfrac{1}{\sizesampos{\nvpos}}
					\outsumop{i=1}{\sizesampos{\nvpos}}{\testres_{i}},
			\end{align}
			with its variance being
			\begin{align}\label{eq_Var_estdenstinf}
				\Var\bbrlv{2}{\estdenstinf(\nvpos)}
				&= \outfrac{\denstinf(\nvpos) \bbrlv{2}{\denstpop(\nvpos) - \denstinf(\nvpos)}}
					{\sizesampos{\nvpos}}.
			\end{align}
			\comment{Then consider the first stage in which
			our goal is to approximate $\totalinf$ by a sample mean.}
			Hence,
			in order to get the estimations of
			$\denstinf(\rvpos_{1}), \omd, \denstinf(\rvpos_{\sizepos})$,
			we have to determine the sample sizes at $\rvpos_{1}, \omd, \rvpos_{\sizepos}$.
			Suppose the expected total sample size is $\sizesam$,
			and $\alloc$ is a positive function on $\region$ such that
			\begin{align}\label{eq_alloc_def}
				\int\nolimits_{\region} \alloc(\nvpos) \denstsam(\nvpos) \diff \nvpos
				= \outfrac{\sizesam}{\sizepos}.
			\end{align}
			When the position $\nvpos \in \region$ is selected in the first stage,
			the sample size at $\nvpos$ will be $\alloc(\nvpos)$, approximately.
			Hereafter, we call $\alloc$ the allocation function of the sample sizes.
			\comment{By some easy calculations, the expectation of the total sample size is just $\sizesam$.}
			Combining the results of the two stages,
			we obtain an unbiased estimator of the total number of infections $\totalinf$,
			\begin{align}\label{eq_esttotalinf}
				\esttotalinf
				&:= \outfrac{1}{\sizepos}
					\outsumop{i=1}{\sizepos}{\outfrac{\estdenstinf(\rvpos_{i})}{\denstsam(\rvpos_{i})}}.
			\end{align}
			The unbiasedness of $\esttotalinf$ can be easily verified
			using the law of total expectation as follows:
			\begin{align}\label{eq_Exp_esttotalinf}
				\Exp\br{\esttotalinf}
				&= \Exp\bbrlv{2}{ \outfrac{\estdenstinf(\rvpos_{1})}{\denstsam(\rvpos_{1})} }
				= \Exp\bbrlv{3}{ \Exp\bbrlv{2}{
					\left. \outfrac{\estdenstinf(\rvpos_{1})}{\denstsam(\rvpos_{1})} \right| \rvpos_{1}
					}}
				= \Exp\bbrlv{2}{ \outfrac{\denstinf(\rvpos_{1})}{\denstsam(\rvpos_{1})} }
				= \int\nolimits_{\region} \outfrac{\denstinf(\nvpos)}{\denstsam(\nvpos)} \cdot \denstsam(\nvpos) \diff \nvpos
				= \totalinf.
			\end{align}
			Similarly, we can obtain its variance
			\begin{align}\label{eq_Var_esttotalinf}
				\Var\br{\esttotalinf}
				&= \outfrac{1}{\sizepos} \bbrlv{3}{
					\Var \bbrlv{2}{ \outfrac{\denstinf(\rvpos_{1})}{\denstsam(\rvpos_{1})} }
					+ \Exp \bbrlv{2}{
					\outfrac{\denstinf(\rvpos_{1}) \bbrlv{2}{\denstpop(\rvpos_{1}) - \denstinf(\rvpos_{1})}}
					{\alloc(\rvpos_{1}) \bbrlv{2}{\denstsam(\rvpos_{1})}^{2}} }
					}
				= \outfrac{1}{\sizepos} \cdot \bbrlv{2}{ v_{0}(\esttotalinf) + v_{1}(\esttotalinf) }.
			\end{align}
			By the central limit theorem,
			\begin{align}\label{eq_CLT_esttotalinf}
				\outfrac{\esttotalinf - \totalinf}{\sqrt{\Var(\esttotalinf)}}
				\stackrel{\mathrm{d}}{\rightarrow} \DistrN(0,1)
				\quad \text{as} \quad \sizepos \rightarrow +\infty,
			\end{align}
			where `$\stackrel{\mathrm{d}}{\rightarrow}$' means the convergence in distribution,
			and $\DistrN(0,1)$ is the standard normal distribution.
			Based on \refeqt{eq_CLT_esttotalinf},
			we can construct the approximate confidence intervals for $\totalinf$
			when $\sizepos$ is large enough.
			It is hinted by the Supplementary Material 
that the least $\sizepos$ required for well approximating this asymptotic distribution
			can be very small, which is completely achievable in practice.
			
		\subsection{Global Likelihood Sampling}\label{sect_prelim_GLS}
		
			The sampling strategy introduced in \refsect{subsect_prelim_basic}
			involves sampling from a bi-variate probability density function $\denstsam$.
			Since the form of $\denstsam$ can be various, multimodal and complicated,
			the extraction of $\rvpos_{1}, \omd, \rvpos_{\sizepos}$ is not easy,
			and some well-known methods, e.g. MCMC and SIR, may be not suitable.
			Instead,
			we adopt the GLS algorithm,
			which applies to the multimodal and complicated cases,
			to generate the $\sizepos$ sampling positions.
			Detailed discussion can refer to \cite{2021_Zhou_HEP}.
			Algorithm \ref{algo_GLS} describes the GLS method for
			generating $\IID$ samples $\rvpos_{1}, \omd, \rvpos_{\sizepos}$ from $\denstsam$.
			The GLS used here is simplified compared to
			the original one in \cite{2015_Wang_conf}.
			
			\begin{algorithm}[!t]
				\caption{GLS algorithm}\label{algo_GLS}
				
				Let the smallest rectangle containing $\region$ be
				$\bar{\region} \subseteq \setR^{2}$;
				without loss of generality, suppose $\bar{\region} = [0,1]^{2}$.
				
				\textbf{Input: }
				the kernel of the sampling density: $\estdenstinfrough$;
				a uniform design on $\bar{\region}$: $\desUD =\setdef{\entUD_{j}}{j \in \nns{\sizeUD}}$.
				
				\begin{stepparenumerate}
					\item \textbf{Loop}.
						Repeat Steps \ref{algo_GLS_shift} to \ref{algo_GLS_sampling}
						for $i = 1, \omd, \sizepos$.
					\item\label{algo_GLS_shift} \textbf{Random shift}.
						Generate $\rvshiftbat{i} \sim \DistrU(\bar{\region})$
						and let
						$\desUDbat{i} = \setdef{\entUD_{j} \oplus \rvshiftbat{i}}{j \in \nns{\sizeUD}}
						\inmathbreaklinespace
						= \desUD \oplus \rvshiftbat{i}$,
						where the operator $\oplus$ means the addition modulo $1$,
						i.e. adding first and then taking the fractional part for each component.
					\item \textbf{Likelihood}.
						For each $\nvpos \in \desUDbat{i}$, let
						$\weightsamsambat{i}(\nvpos) = \infrac{\estdenstinfrough(\nvpos)}
						{\insumop{\vecs{y} \in \desUDbat{i}}{}{\estdenstinfrough(\vecs{y})}}$,
						where $\estdenstinfrough(\nvpos) = 0$
						for $\nvpos \in \bar{\region} \setminus \region$.
						Then
						$\distrsamsambat{i} =
						\setdef{(\nvpos, \weightsamsambat{i}(\nvpos))}{\nvpos \in \desUDbat{i}}$
						is a multinomial distribution on $\desUDbat{i}$.
					\item\label{algo_GLS_sampling} \textbf{Sampling}.
						Generate $\rvpos_{i}$ in $\desUDbat{i}$
						from the multinomial distribution $\distrsamsambat{i}$.
				\end{stepparenumerate}
				
				\textbf{Output: }
				$\sizepos$ $\IID$ samples from $\denstsam$: $\rvpos_{1}, \omd, \rvpos_{\sizepos}$.
			\end{algorithm}
			
			The performance of Algorithm \ref{algo_GLS} depends on the uniformity of $\desUD$.
			Intuitively,
			the better the uniformity of $\desUD$,
			the better $\distrsamsambat{i}$ approximates $\denstsam$ for each $i \in \nns{\sizepos}$,
			and, therefore, the better
			the empirical distribution of $\rvpos_{1}, \omd, \rvpos_{\sizepos}$ approximates $\denstsam$.
			There are some additional remarks about
			the region $\bar{\region}$ and the uniform design $\desUD$ in Algorithm \ref{algo_GLS} in the Supplementary Material.
			
	\section{Optimal Settings of the Parameters}\label{sect_optimal}
		
		In the two-stage sampling strategy introduced in \refsect{subsect_prelim_basic},
		there are three adjustable parameters:
		the sampling density $\denstsam$,
		the allocation function of the sample sizes $\alloc$,
		and the number of sampling positions $\sizepos$.
		From \refeqt{eq_Exp_esttotalinf},
		\refeqt{eq_Var_esttotalinf} and \refeqt{eq_CLT_esttotalinf},
		these three parameters do not affect the unbiasedness of $\esttotalinf$,
		but do affect its variance and distribution.
		In this section, we discuss the optimal settings of these parameters,
		where `optimal' means to minim\UKUS{is}e the variance of $\esttotalinf$.

		First, we consider $\alloc$,
		since it only affects the term $v_{1}(\esttotalinf)$ in \refeqt{eq_Var_esttotalinf}.
		By the Lagrange multiplier method, we can find that
		under the constraint \eqref{eq_alloc_def},
		$v_{1}(\esttotalinf)$ is minim\UKUS{is}ed when
		for any $\nvpos \in \region$,
		\begin{align}\label{eq_alloc_optimal}
			\alloc(\nvpos)
			&= \outfrac{\sizesam }{\sizepos \int\nolimits_{\region} \denstvarker{(\vecs{y})} \diff \vecs{y}}
				\cdot \outfrac{\denstvarker{(\nvpos)}}{\denstsam(\nvpos)},
		\end{align}
		where $\denstvarker{} := \sqrt{\denstinf (\denstpop-\denstinf)}$.
		The minimum of $\infrac{v_{1}(\esttotalinf)}{\sizepos}$ is
		$\infrac{\bbrlv{2}{\int\nolimits_{\region} \denstvarker{(\nvpos)} \diff \nvpos}^{2}}{\sizesam}$,
		which is independent of the settings of $\denstsam$ and $\sizepos$.
		Then note that $v_{0}(\esttotalinf)=0$ when $\denstsam \propto \denstinf$,
		hence no matter what $\sizepos$ is,
		when $\denstsam \propto \denstinf$ and $\alloc$ is set as \refeqt{eq_alloc_optimal},
		$\Var(\esttotalinf)$ is minim\UKUS{is}ed with the minimum being
		\begin{align}\label{eq_Var_esttotalinf_min}
			\outfrac{1}{\sizesam}
			\bbrlv{2}{ \int\nolimits_{\region} \denstvarker{(\nvpos)} \diff \nvpos }^{2}.
		\end{align}

		However,
		the above optimal settings of $\denstsam$ and $\alloc$ cannot be achieved in practice
		because those optimal settings depend on $\denstinf$,
		which is unknown, and is just what we want to estimate.
		Instead,
		we need to find the nearly optimal settings of $\denstsam$ and $\alloc$.
		We first consider the following mechanism to determine an initial rough estimate of $\denstinf$.
		Without any extra knowledge,
		the only information about $\denstinf$ is
		$0 \leqslant \denstdiag \leqslant \denstinf \leqslant \denstpop$,
		which implies that there exists a function $\coefdenstinf$ from $\region$ to $[0,1]$
		such that $\denstinf = \coefdenstinf \cdot \denstpop + (1-\coefdenstinf) \cdot \denstdiag$.
		It is impossible to know the form of $\coefdenstinf$ unless some extra information is given.
		A simple but reasonable way is to choose a proper constant $\estcoefdenstinf \in [0,1]$
		and take
		\begin{align}\label{eq_estdenstinfrough}
			\estdenstinfrough := \estcoefdenstinf \cdot \denstpop + (1 - \estcoefdenstinf) \cdot \denstdiag
		\end{align}
		as an initial rough estimate of $\denstinf$.
		This estimation combines the information of both $\denstpop$ and $\denstdiag$,
		which can make our method efficient.
		With the estimator $\estdenstinfrough$, the nearly optimal setting of $\denstsam$ is
		${\estdenstinfrough}/
		{\bbrlv{2}{\estcoefdenstinf \cdot \totalpop + (1-\estcoefdenstinf) \cdot \totaldiag}}$.
		As for $\alloc$, we can use
		$\estdenstvarkerrough{} := \sqrt{\estdenstinfrough (\denstpop-\estdenstinfrough)}$
		to estimate $\denstvarker{}$ in \refeqt{eq_alloc_optimal},
		but there remains an integral
		$\int\nolimits_{\region} \denstvarker{(\vecs{y})} \diff \vecs{y}$ to approximate.
		Since the form of $\denstvarker{}$ is generally complicated,
		it is appropriate to approximate this integral using the Monte Carlo method.
		In fact,
		since the $\sizepos$ positions
		$\rvpos_{1}, \omd, \rvpos_{\sizepos}$ are $\IID$ samples from $\denstsam$,
		the sample mean
		\begin{align*}
			\outfrac{1}{\sizepos} \cdot \outsumop{i=1}{\sizepos}{
				\outfrac{\estdenstvarkerrough{(\rvpos_{i})}}{\denstsam(\rvpos_{i})}
			}
		\end{align*}
		is an unbiased estimator of the integral
		$\int\nolimits_{\region} \denstvarker{(\vecs{y})} \diff \vecs{y}$.
		Therefore, for each $i \in \nns{\sizepos}$,
		the nearly optimal sample size at the sampling position $\rvpos_{i}$,
		which is an approximation of the exact optimal $\alloc(\rvpos_{i})$,
		is
		\begin{align*}
			\sizesampos{\rvpos_{i}}
			:= \outfrac{\estdenstvarkerrough{(\rvpos_{i})} / \denstsam(\rvpos_{i})}
				{\outsumop{j=1}{\sizepos}{\estdenstvarkerrough{(\rvpos_{j})} / \denstsam(\rvpos_{j})}}
				\cdot \sizesam.
		\end{align*}
		With the nearly optimal setting of $\denstsam$,
		it is simplified to
		\begin{align}\label{eq_sizesampos_opt}
			\sizesampos{\rvpos_{i}}
			= \outfrac{\sqrt{\bbrlv{2}{\denstpop(\rvpos_{i})-\estdenstinfrough(\rvpos_{i})} /
				\estdenstinfrough(\rvpos_{i})}}
				{\outsumop{j=1}{\sizepos}{
					\sqrt{\bbrlv{2}{\denstpop(\rvpos_{j})-\estdenstinfrough(\rvpos_{j})} /
					\estdenstinfrough(\rvpos_{j})}
				}}
				\cdot \sizesam.
		\end{align}
		This allocation method of sample sizes has the benefit that
		the total sample size is exactly $\sizesam$,
		which meets the requirement in most practical situations.

		Next, we consider the number of sampling positions $\sizepos$.
		From \refeqt{eq_alloc_def} and \eqref{eq_alloc_optimal},
		$\alloc$ is approximately proportional to $\infrac{1}{\sizepos}$,
		thus $\infrac{v_{1}(\esttotalinf)}{\sizepos}$ in \refeqt{eq_Var_esttotalinf}
		is always approximately independent of $\sizepos$,
		and a large $\sizepos$ can reduce $\Var(\esttotalinf)$
		by reducing $\infrac{v_{0}(\esttotalinf)}{\sizepos}$.
		Therefore,
		$\sizepos$ should be large in order to reduce the variance
		when $\denstsam$ is not exactly proportional to $\denstinf$,
		as long as $\sizesampos{\rvpos_{i}}$
		is not too small to estimate the value of $\denstinf$ at $\rvpos_{i}$
		for each $i \in \nns{\sizepos}$.
		By \eqref{eq_CLT_esttotalinf},
		a large $\sizepos$ also helps to obtain a good approximate distribution of $\esttotalinf$.
		The numerical simulations in the Supplementary Material 
		show that larger $\sizepos$ can notably reduce the variance
		and improve the coverage rate of the confidence interval.
		Note that in practice,
		larger $\sizepos$ may also increase the cost and the difficulty of the sampling survey,
		since more positions have to be sampled.
		These results are consistent with the classical sampling theory \citep{1977_Cochran_Wiley}.
		
		In addition,
		the costs at different sampling positions may be different in practice,
		which can be quantified by a cost function $\cost$ over $\region$.
		Then our goal is to find the appropriate parameters
		which can minim\UKUS{is}e both of $\Var(\esttotalinf)$ and the total cost.
		For example,
		the weighted sum of $\Var(\esttotalinf)$ and the total cost
		can be used as an optimality criterion, i.e.,
		$\Var(\esttotalinf) + \inmathbreaklinespace
		\coefcost \int_{\region} \cost(\nvpos)\alloc(\nvpos)\denstsam(\nvpos) \diff\nvpos$
		where $\coefcost \in [0,+\infty)$ reflects the importance of the total cost.
		For such cases,
		an analysis similar to the above can be performed,
		but it may be difficult to derive the explicit expressions.
		Instead,
		the numerical optim\UKUS{is}ation algorithms can be considered to solve this problem,
		which is beyond the scope of this paper.
		
	\section{Sampling and Estimating}\label{sect_prac}

		Based on the discussion about the nearly optimal settings of the parameters
		in \refsect{sect_optimal},
		we show the complete sampling strategy and the estimating method in this section.
		Some suggestions for the practical implementation are also given.
		
		
			The details of the two-stage sampling strategy are described in Algorithm \ref{algo_sample},
			which combines the basic approach in \refsect{subsect_prelim_basic}
			with the GLS algorithm in \refsect{sect_prelim_GLS}.
			
			\begin{algorithm}
				\caption{Two-stage sampling strategy}\label{algo_sample}
				
				\textbf{Input: }
				a proper constant in $[0,1]$ to give a rough estimator of $\denstinf$: $\estcoefdenstinf$;
				the uniform design used in Algorithm \ref{algo_GLS}: $\desUD$;
				the total sample size: $\sizesam$;
				the number of sampling positions: $\sizepos$.
				
				\begin{stepparenumerate}
					\item \textbf{Rough estimate}.
						Obtain $\estdenstinfrough$,
						an initial rough estimator of $\denstinf$,
						by \refeqt{eq_estdenstinfrough}.
						The kernel of the sampling density $\denstsam$ is set to be $\estdenstinfrough$.
					\item\label{algo_sample_position} \textbf{Sampling positions}.
						Generate $\sizepos$ $\IID$ sampling positions
						$\rvpos_{1}, \omd, \rvpos_{\sizepos}$ in $\region$
						from the sampling density $\denstsam$
						by the GLS in Algorithm \ref{algo_GLS}.
					\item\label{algo_sample_alloc} \textbf{Allocation of sample sizes}.
						For each $i \in \nns{\sizepos}$,
						calculate $\round(\sizesampos{\rvpos_{i}})$,
						the sample size at the sampling position $\rvpos_{i}$,
						where $\sizesampos{\rvpos_{i}}$ is calculated by \refeqt{eq_sizesampos_opt}
						and `$\round$' is the function
						rounding a real number into the nearest integer.
					\item \textbf{Test}.
						For each $i \in \nns{\sizepos}$,
						select $\round(\sizesampos{\rvpos_{i}})$ people at position $\rvpos_{i}$ by
						the simple random sampling
						and then implement tests on them.
				\end{stepparenumerate}
			\end{algorithm}
			
			Next, we give some remarks about Algorithm \ref{algo_sample} as follows.
			
			\begin{enumerate}
				\item About $\sizesam$.
					In practice, the $\sizesam$ is usually determined
					by both the requirement of precision and the restriction of costs,
					thus $\Var(\esttotalinf)$ should be roughly estimated
					before the implementation of the sampling strategy.
					This can be done by using \refeqt{eq_Var_esttotalinf_min},
					which can be approximated by
					\begin{align*}
						\outfrac{1}{\sizesam} \bbrlv{2}{ \outfrac{\measure(\region)}{\sizeUD}
						\outsumop{\nvpos \in D}{}{ \estdenstvarkerrough{(\nvpos)}} }^{2}
					\end{align*}
					where
					\comment{$\estdenstvarkerrough{} = \sqrt{\estdenstinfrough (\denstpop-\estdenstinfrough)}$
					and}
					$\measure(\region)$ is the area of $\region$.
					The $\sizesam$ should be set
					such that the above estimator of $\Var(\esttotalinf)$ is
					smaller than some pre-defined threshold about the precision.
				\item About $\sizesampos{\rvpos_{i}}$.
					As mentioned in \refsect{sect_optimal},
					$\sizepos$ should be as large as possible,
					provided that the cost will not exceed the budget,
					and no $\sizesampos{\rvpos_{i}}$ is too small to estimate $\denstinf(\rvpos_{i})$.
					\comment{
					But if some $\sizesampos{\rvpos_{i}}$'s appear too small,
					we can remove those sampling positions with
					small sample sizes there,
					reduce $\sizepos$,
					and then allocate $\sizesam$ into the remaining sampling positions
					according to \refeqt{eq_sizesampos_opt}.
					\comment{
					the least sample size there,
					reduce the number of sampling positions $\sizepos$ by $1$,
					and then allocate $\sizesam$ into the remaining sampling positions
					according to \refeqt{eq_sizesampos_opt}.
					This operation can be repeated until
					none of the sample sizes at those sampling positions is too small.
					}
					Another way to avoid this situation is
					}
					An additional way to avoid small $\sizesampos{\rvpos_{i}}$ is
					to modify the calculation method of the sample sizes at those sampling positions,
					i.e., the $\sizesampos{\rvpos_{i}}$
					in Step \ref{algo_sample_alloc} of Algorithm \ref{algo_sample} as
					\begin{align}\label{eq_sizesampos_opt_mod}
						\sizesampos{\rvpos_{i}}
						:= \outfrac{
							\sqrt{\bbrlv{2}{\denstpop(\rvpos_{i})-\estdenstinfrough(\rvpos_{i})} /
							\estdenstinfrough(\rvpos_{i})}}
							{\outsumop{j=1}{\sizepos}{
							\sqrt{\bbrlv{2}{\denstpop(\rvpos_{j})-\estdenstinfrough(\rvpos_{j})} /
							\estdenstinfrough(\rvpos_{j})}
							}}
							\cdot (1-\coefsizesampos) \sizesam
							+ \outfrac{1}{\sizepos} \cdot \coefsizesampos \sizesam,
					\end{align}
					where $\coefsizesampos \in [0,1]$ is a properly chosen constant.
					Since the allocation method of sample sizes in \refeqt{eq_sizesampos_opt} is nearly optimal,
					$\coefsizesampos$ in \refeqt{eq_sizesampos_opt_mod} should not be too large.
					The proper setting of $\coefsizesampos$ should
					make the minimum sample size among those sampling positions
					just achieve some pre-defined threshold.
					\comment{of the sample size at a single sampling position.}
					Further,
					if $\sizesampos{\rvpos_{i}}$ is not very small
					compared to the population around the sampling position $\rvpos_{i}$,
					finite population corrections
					\citep{1977_Cochran_Wiley, 2019_Lohr_BrooksCole}
					should be applied.
					For such case,
					expression \refeqt{eq_Var_estdenstinf} becomes
					\begin{align*}
						\Var\bbrlv{2}{\estdenstinf(\nvpos)}
						&= \bbrlv{2}{1-\ratesam(\nvpos)} \cdot
							\outfrac{\denstinf(\nvpos) \bbrlv{2}{\denstpop(\nvpos) - \denstinf(\nvpos)}}
							{\sizesampos{\nvpos}}
					\end{align*}
					where $\ratesam(\nvpos)$ refers to the sampling fraction at $\nvpos \in \region$,
					and $v_{1}(\esttotalinf)$ in \refeqt{eq_Var_esttotalinf} becomes
					\begin{align*}
						\Exp \bbrlv{3}{
						\bbrlv{2}{1-\ratesam(\rvpos_{1})} \cdot
						\outfrac{\denstinf(\rvpos_{1}) \bbrlv{2}{\denstpop(\rvpos_{1}) - \denstinf(\rvpos_{1})}}
						{\alloc(\rvpos_{1}) \bbrlv{2}{\denstsam(\rvpos_{1})}^{2}}
						}.
					\end{align*}
				\item About the sampling survey at each sampling position.
					For each $i \in \nns{\sizepos}$, in practice,
					the $\sizesampos{\rvpos_{i}}$ samples come from not exactly $\rvpos_{i}$,
					but a neighb\UKUS{our}hood of $\rvpos_{i}$.
					For convenience,
					this neighb\UKUS{our}hood can be chosen as some small district,
					like a city, village, or community,
					containing $\rvpos_{i}$ or just close to $\rvpos_{i}$.
					It won't affect the property of this sampling strategy
					as long as the diameter of the neighb\UKUS{our}hood is negligible
					compared to the whole region $\region$.
					For example,
					the samples at each $\rvpos_{i}$ can be drawn from
					the people whose current residences are within
					the circle cent\UKUS{red} at $\rvpos_{i}$ with radius $10 \mathrm{km}$.
					Some other popular sampling methods,
					such as the stratified sampling, cluster sampling,
					multi-stage sampling and other techniques \citep{2019_Lohr_BrooksCole},
					can be used to estimate $\denstinf(\rvpos_{i})$ more efficiently.
					The corresponding results can be obtained similarly with suitable modifications.
			\end{enumerate}
			
			In addition,
			it is difficult to theoretically optim\UKUS{is}e the setting of $\estcoefdenstinf$,
			but the numerical studies will give some recommendations in \refsect{sect_simu_coef}.
			
		
			Our main purpose is to estimate the cumulative total number of infections $\totalinf$.
			With the testing results obtained by the above sampling strategy,
			we can estimate $\totalinf$
			by using \refeqttageqts \eqref{eq_estdenstinf} and \eqref{eq_esttotalinf}.
			The variance of the estimator can also be estimated by \refeqt{eq_Var_esttotalinf},
			and then an approximate confidence interval (CI)
			can be constructed according to \eqref{eq_CLT_esttotalinf}.
			This estimating procedure is discribed in detail as follows.
			
			\begin{stepparenumerate}
				\item For $i = 1, \omd, \sizepos$,
					calculate $\estdenstinf(\rvpos_{i})$ by \refeqt{eq_estdenstinf}
					using the testing results at the sampling position $\rvpos_{i}$,
					and calculate $\denstsam(\rvpos_{i})$
					where
					$\denstsam = {\estdenstinfrough}
					/{\bbrlv{2}{\estcoefdenstinf \cdot \totalpop + (1-\estcoefdenstinf) \cdot \totaldiag}}$.
				\item Obtain the point estimator $\esttotalinf$ of
					the total number of infections $\totalinf$
					according to \refeqt{eq_esttotalinf}.
				\item\label{algo_est_var}
					According to \refeqt{eq_Var_esttotalinf}, let
					\begin{align}
					\begin{aligned}\label{eq_est_Var_esttotalinf}
						\hat{v}_{0}(\esttotalinf)
						&:= \outfrac{1}{\sizepos-1}
							\outsumop{i=1}{\sizepos}{\bbrlv{2}{
							\outfrac{\estdenstinf(\rvpos_{i})}{\denstsam(\rvpos_{i})}
							- \esttotalinf
							}^{2}}, \\
						\hat{v}_{1}(\esttotalinf)
						&:= \outfrac{1}{\sizepos} \outsumop{i=1}{\sizepos}{
							\bbrlv{2}{1-\ratesam(\rvpos_{i})} \cdot
							\outfrac{\estdenstinf(\rvpos_{i})
							\bbrlv{2}{\denstpop(\rvpos_{i}) - \estdenstinf(\rvpos_{i})}}
							{\sizesampos{\rvpos_{i}} \bbrlv{2}{\denstsam(\rvpos_{i})}^{2}}}, \\
						\hat{v}(\esttotalinf)
						&:= \outfrac{1}{\sizepos} \cdot
							\bbrlv{2}{\hat{v}_{0}(\esttotalinf) + \hat{v}_{1}(\esttotalinf)},
					\end{aligned}
					\end{align}
					then $\hat{v}(\esttotalinf)$ is an estimator of $\Var(\esttotalinf)$.
				\item According to \eqref{eq_CLT_esttotalinf},
					an approximate $1-\alpha$ CI of $\totalinf$ is
					$\big[ \esttotalinf - z_{\alpha/2} \sqrt{\hat{v}(\esttotalinf)}, \inmathbreaklinespace
					\esttotalinf + z_{\alpha/2} \sqrt{\hat{v}(\esttotalinf)} \big]$,
					where $z_{\alpha/2}$ is the upper $\alpha/2$-quantile of $\DistrN(0,1)$.
			\end{stepparenumerate}
			
\begin{figure}[!t]
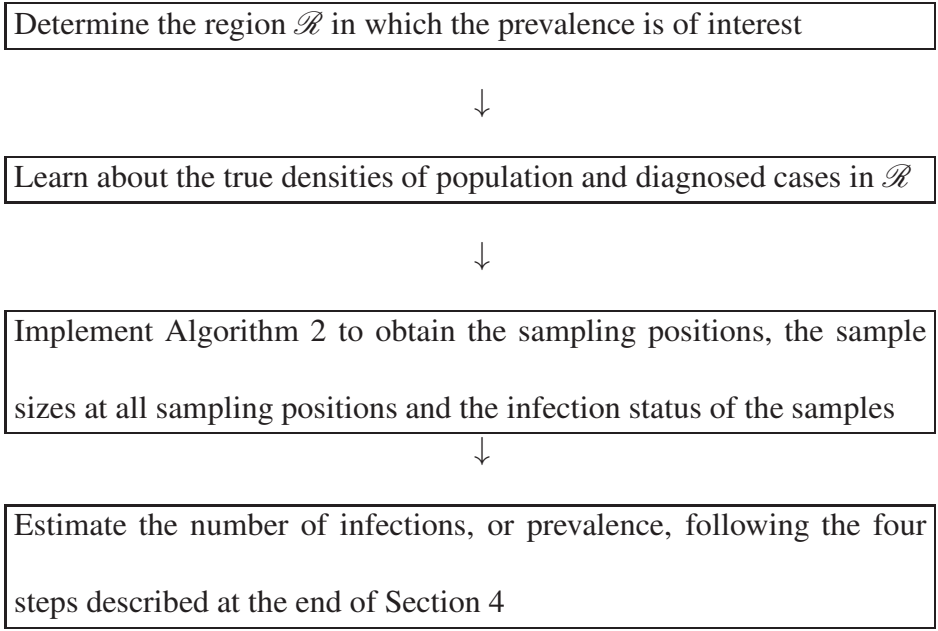

			\centering
			\begin{LZflowdiagram}
				\flowstep{Determine the region $\region$ in which the prevalence is of interest}
				\flowstep{Learn about
					the true densities of population and diagnosed cases in $\region$}
				\flowstep{Implement Algorithm \ref{algo_sample} to obtain
					the sampling positions, the sample sizes at all sampling positions
					and the infection status of the samples}
				\flowstep{Estimate the number of infections, or prevalence,
					following the four steps described at the end of \refsect{sect_prac}}
			\end{LZflowdiagram}
			\caption{The complete procedure for implementing the proposed sampling method.}
			\label{fig_prac_proc}
		\end{figure}

			Hence,
			by the two-stage sampling strategy and the estimating method above,
			we can obtain the estimator of $\totalinf$, its estimated variance and approximate CI.
			The complete procedure is summar\UKUS{is}ed in \reffig{fig_prac_proc}.
			In the next two sections,
			we will show some numerical simulations and a practical example
			to verify the validity of our proposed sampling strategy.
			

	\section{Numerical Simulation}\label{sect_simu}
		
		In \refsect{sect_optimal} and \refsect{sect_prac},
		we discussed
		how to set the parameters in the two-stage sampling strategy,
		except for the coefficient $\estcoefdenstinf$ in \refeqt{eq_estdenstinfrough}.
		In this section,
		we first give a robust setting for $\estcoefdenstinf$ in the minimax sense
		through some numerical simulations,
		when there is little information about the underlying true $\denstinf$.
		Then we compare our proposed sampling strategy with some other popular methods
		to verify the efficiency of our method.
		
		\subsection{Robust Setting of $\estcoefdenstinf$}\label{sect_simu_coef}
			
			In the oracle situation,
			when the exact optimal settings of $\alloc$ and $\denstsam$ discussed in \refsect{sect_optimal}
			can be achieved,
			the variance of $\esttotalinf$ is minim\UKUS{is}ed,
			denoted by $\Var(\esttotalinf \mid \denstsam \propto \denstinf)$.
			On the other hand,
			when we do not know the true $\denstinf$
			and have to use the initial rough estimator \refeqt{eq_estdenstinfrough}
			to determine the sample sizes and the sampling density,
			the variance of $\esttotalinf$,
			denoted by $\Var(\esttotalinf \mid \denstsam \propto \estdenstinfrough)$,
			depends on the quality of $\estdenstinfrough$,
			and thus depends on $\estcoefdenstinf$.
			Therefore,
			in order to measure the performance of
			different settings of $\estdenstinfrough$ or $\estcoefdenstinf$,
			we define the standard\UKUS{is}ed standard deviation ($\critSSD$) as
			\begin{align*}
				\critSSD(\estcoefdenstinf)
				= \critSSD(\estdenstinfrough)
				:= \sqrt{\outfrac{\Var(\esttotalinf \mid \denstsam \propto \estdenstinfrough)}
					{\Var(\esttotalinf \mid \denstsam \propto \denstinf)}}.
			\end{align*}
			Our goal is to find the setting of $\estcoefdenstinf$ such that
			the maximum of $\critSSD$ over all possible true $\denstinf$'s
			is minim\UKUS{is}ed at that $\estcoefdenstinf$.
			
			In the simulations,
			the region $\region = [0,1]^{2}$ is the unit square
			and it is divided into four equal-sized sub-squares.
			Let the populations in them be 20, 40, 60 and 80, multiplied by $1 \times 10^{4}$ respectively,
			and the numbers of cases in them be 6, 8, 4 and 2, multiplied by $1 \times 10^{4}$ respectively.
			Dividing those numbers by $1/4$ will obtain the corresponding densities in the sub-squares.
			Following is the corresponding graph.
			\begin{align*}
				\text{Populations ($\times 10^{4}$): }
				\begin{array}{|c|c|}
					\hline
					20 & 40 \\
					\hline
					60 & 80 \\
					\hline
				\end{array}
				\ ,\quad
				\text{Numbers of cases ($\times 10^{4}$): }
				\begin{array}{|c|c|}
					\hline
					6 & 8 \\
					\hline
					4 & 2 \\
					\hline
				\end{array}
				\ .
			\end{align*}
			The total population $\totalpop = 200 \times 10^{4}$
			and the total number of cases $\totaldiag = 20 \times 10^{4}$.
			We set the size of the uniform design used in the GLS algorithm $\sizeUD = 210$,
			the total sample size $\sizesam = 1 \times 10^{4}$ and
			the number of sampling positions $\sizepos = 50$.
			In different groups of simulations,
			the settings of $\denstinf$ will be different.
			For each setting of $\denstinf$ and $\estcoefdenstinf$,
			$\Var(\esttotalinf)$ is calculated using the sample variance of $\esttotalinf$
			over 200 independent simulations.
			
\begin{figure}[!t]
			\centering
			\subfigure[E1]{
				\includegraphics[width=0.3\textwidth]{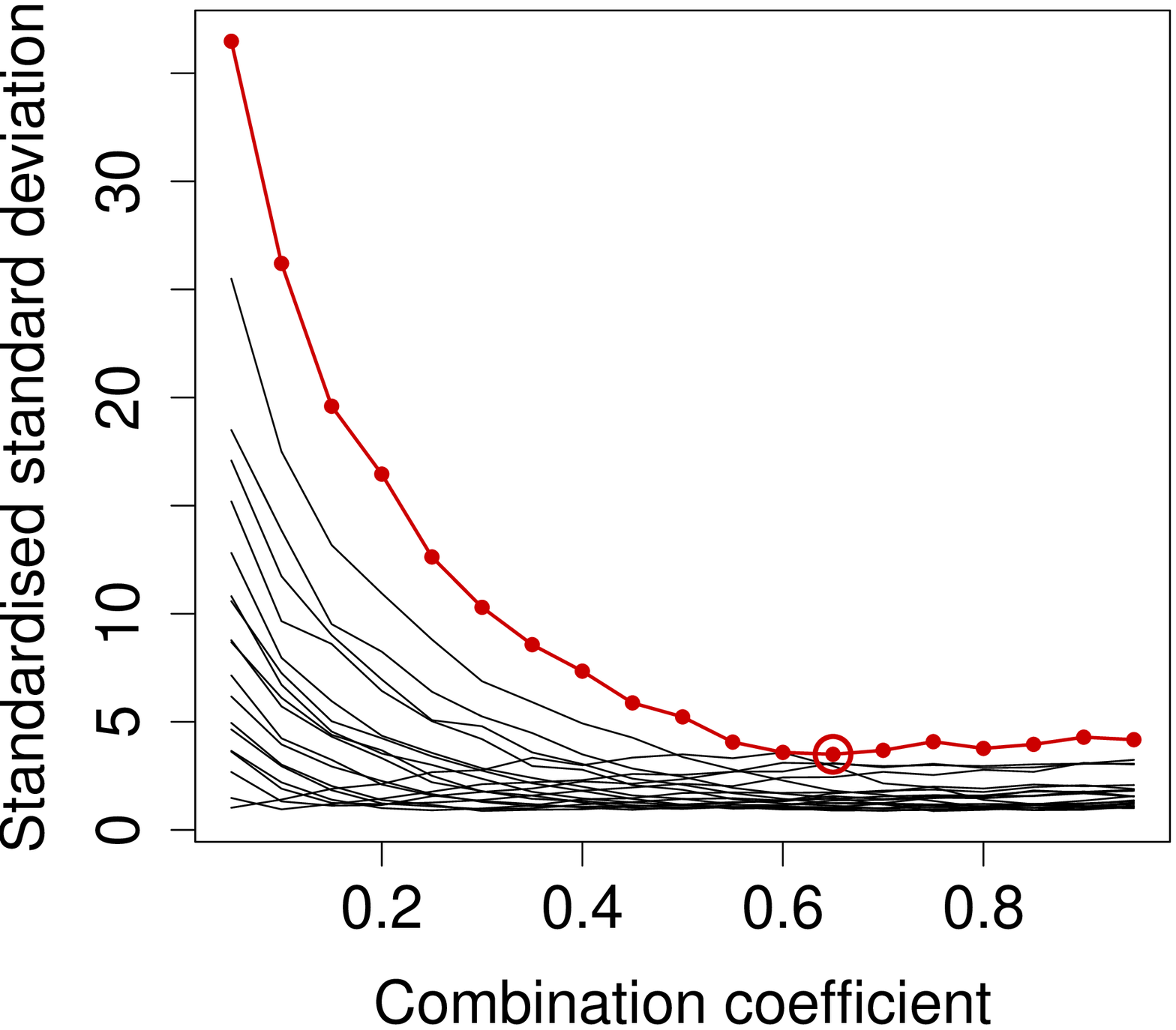}
				\subfiglabel{fig_simu_SSD_E.E1}
			}
			\hspace{-0.05\textwidth}
			\subfigure[E2]{
				\includegraphics[width=0.3\textwidth]{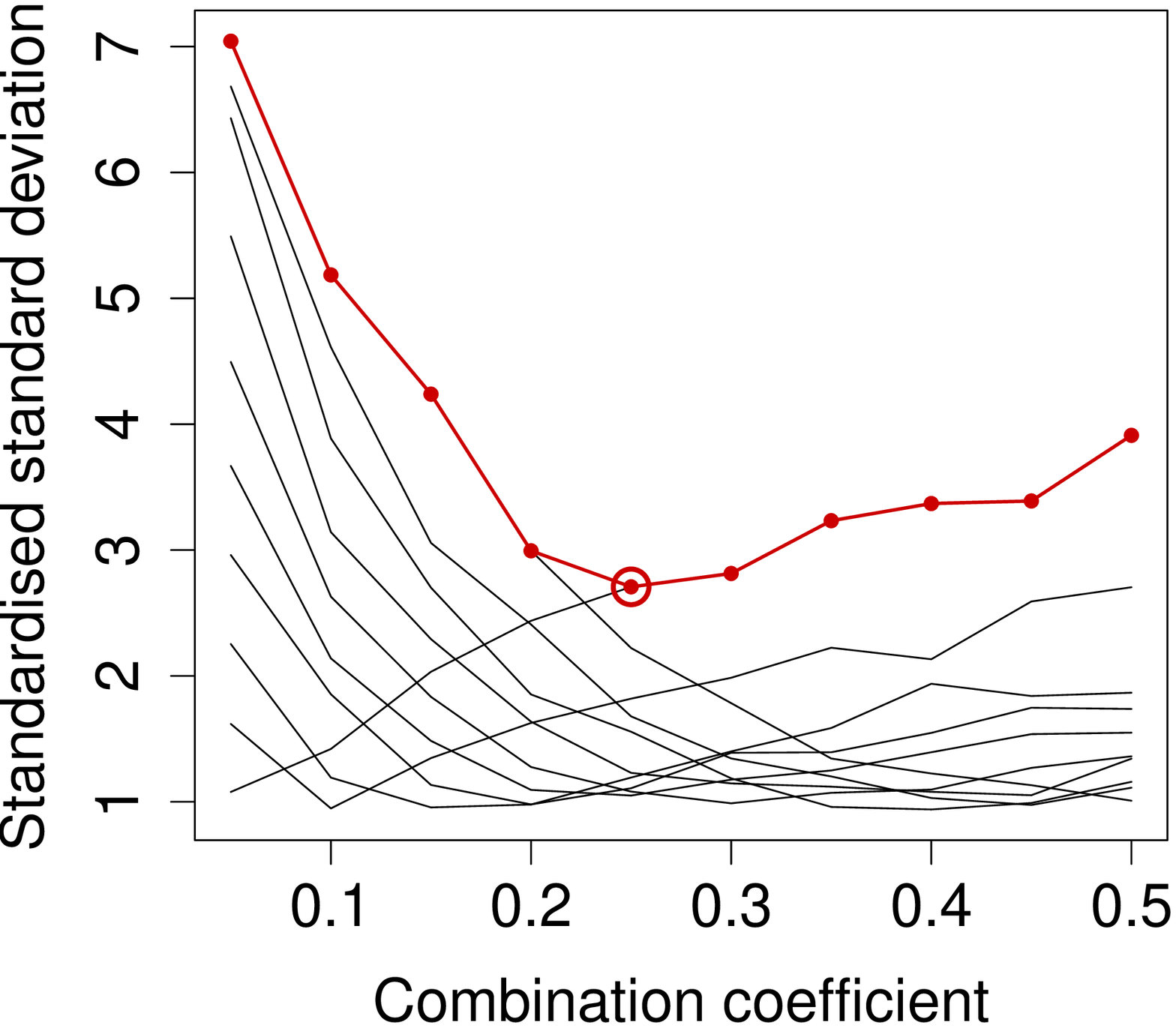}
				\subfiglabel{fig_simu_SSD_E.E2}
			}
			\hspace{-0.05\textwidth}
			\subfigure[E3]{
				\includegraphics[width=0.3\textwidth]{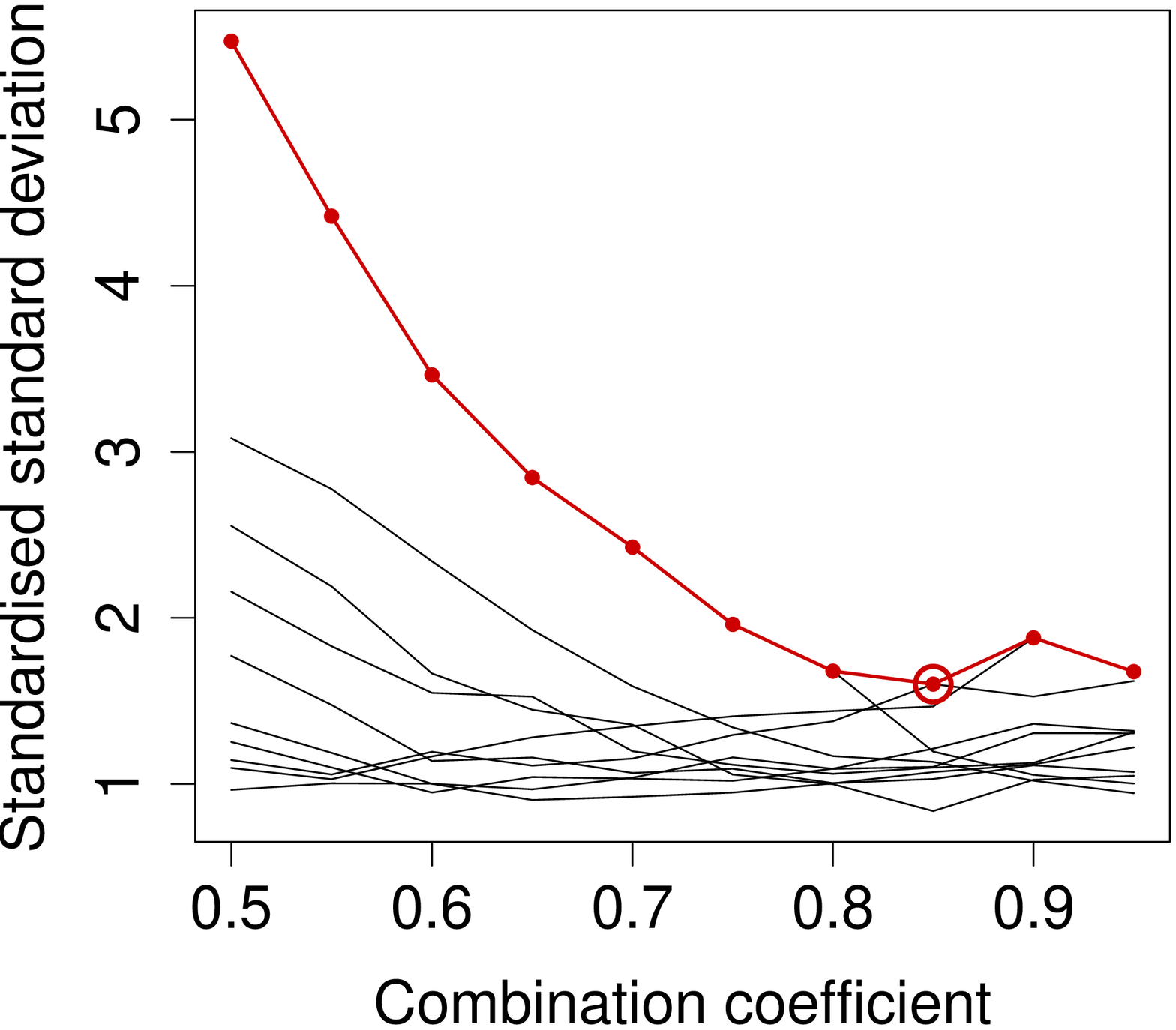}
				\subfiglabel{fig_simu_SSD_E.E3}
			}
			\caption{Results of the simulations in Series E: curves of $\critSSD$ against $\estcoefdenstinf$.}
			\label{fig_simu_SSD_E}
		\end{figure}

			In the first series of simulations, Series E, assume
			the true density of the infections $\denstinf$
			is a convex combination of $\denstpop$ and $\denstdiag$,
			i.e. $\denstinf = \coefdenstinf \denstpop + (1-\coefdenstinf) \denstdiag$,
			where $\coefdenstinf \in [0,1]$ is a constant.
			Series E contains three groups of simulations, E1 to E3,
			whose settings are shown in \reftab{tab_simu_coef_setting_E},
			and the corresponding results are given in \reffig{fig_simu_SSD_E}.
			For each setting of $\denstinf$ (or $\coefdenstinf$) in Group E1,
			the coefficient $\estcoefdenstinf$ takes $19$ different values,
			and the corresponding $19$ values of $\critSSD(\estcoefdenstinf)$
			form a black curve in \reffig{fig_simu_SSD_E.E1}.
			The robust setting of $\estcoefdenstinf$ is
			the one that minim\UKUS{is}es the maximum $\critSSD$ among the $19$ settings of $\denstinf$,
			i.e., the one that minim\UKUS{is}es the red bold curve in \reffig{fig_simu_SSD_E.E1},
			which is marked out by a red circle.
			The other two subfigures are obtained similarly.
			The three subfigures in \reffig{fig_simu_SSD_E} present a similar phenomenon that
			in the domain of $\coefdenstinf$,
			the maximum $\critSSD$ is large
			when $\estcoefdenstinf$ is close to the bounds of the domain,
			while the maximum $\critSSD$ is minim\UKUS{is}ed
			when $\estcoefdenstinf$ is neither too large nor too small.
			Therefore, \reffig{fig_simu_SSD_E} indicates that
			when a possible range of the true value of $\coefdenstinf$ is available,
			the mid-point of that range may be a robust setting of $\estcoefdenstinf$.
			
		\begin{table}[!t]
			\centering
			\caption{Settings of $\denstinf$ and $\estcoefdenstinf$ in the simulations in Series E.}
			\begin{tabular*}{\textwidth}{@{\extracolsep\fill}cccc@{\extracolsep\fill}}
				\toprule
				& Group E1 & Group E2 & Group E3 \\
				\midrule
				$\coefdenstinf$ & $\{0.05, 0.1, \omd, 0.9, 0.95\}$
				& $\{0.05, 0.1, \omd, 0.45, 0.5\}$ & $\{0.5, 0.55, \omd, 0.9, 0.95\}$ \\
				$\estcoefdenstinf$ & $\{0.05, 0.1, \omd, 0.9, 0.95\}$
				& $\{0.05, 0.1, \omd, 0.45, 0.5\}$ & $\{0.5, 0.55, \omd, 0.9, 0.95\}$ \\
				\bottomrule
			\end{tabular*}
			\label{tab_simu_coef_setting_E}
		\end{table}

	
\begin{figure}[!t]
			\centering
			\subfigure[R1]{
				\includegraphics[width=0.3\textwidth]{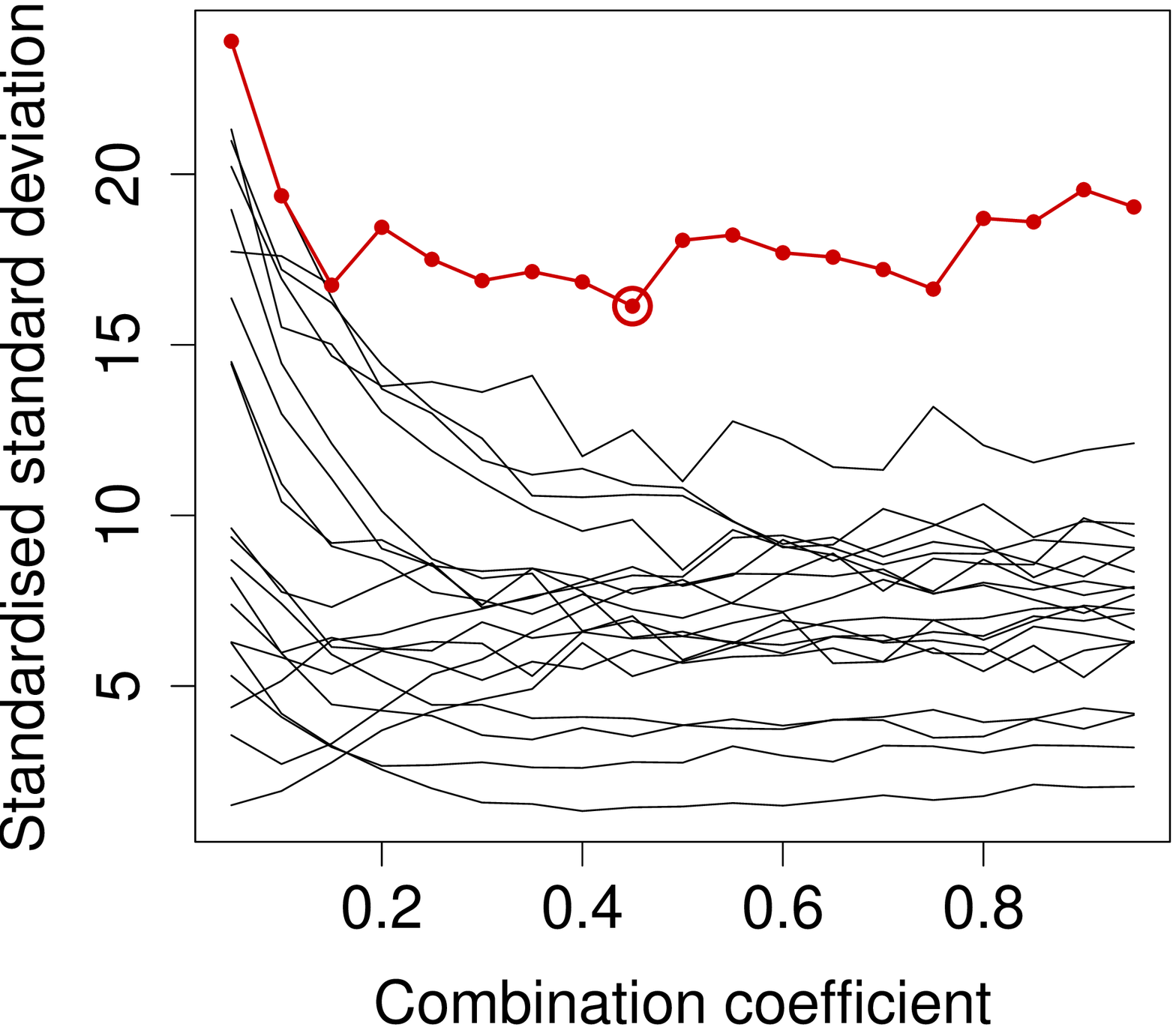}
				\subfiglabel{fig_simu_SSD_R.R1}
			}
			\hspace{-0.05\textwidth}
			\subfigure[R2]{
				\includegraphics[width=0.3\textwidth]{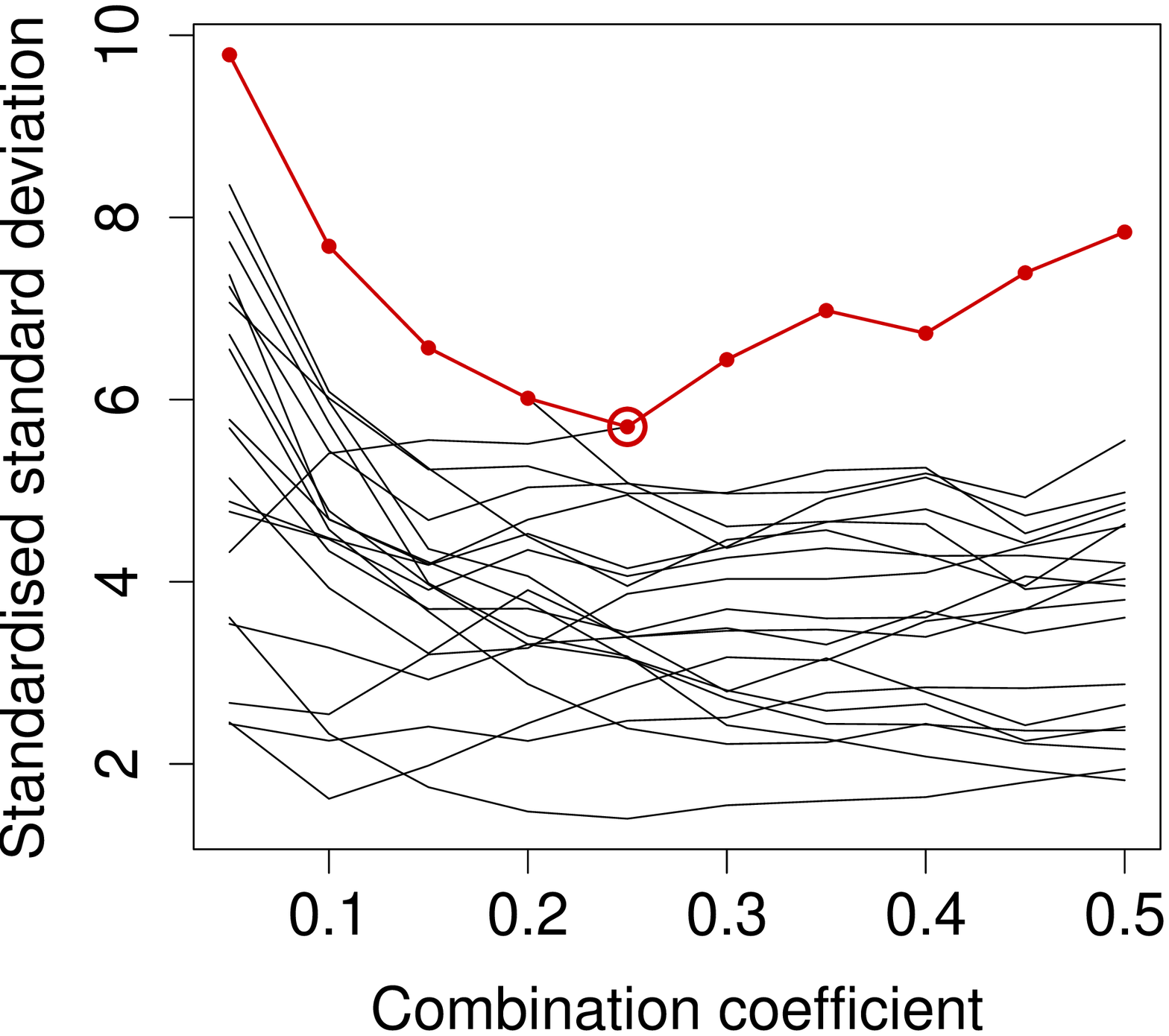}
				\subfiglabel{fig_simu_SSD_R.R2}
			}
			\hspace{-0.05\textwidth}
			\subfigure[R3]{
				\includegraphics[width=0.3\textwidth]{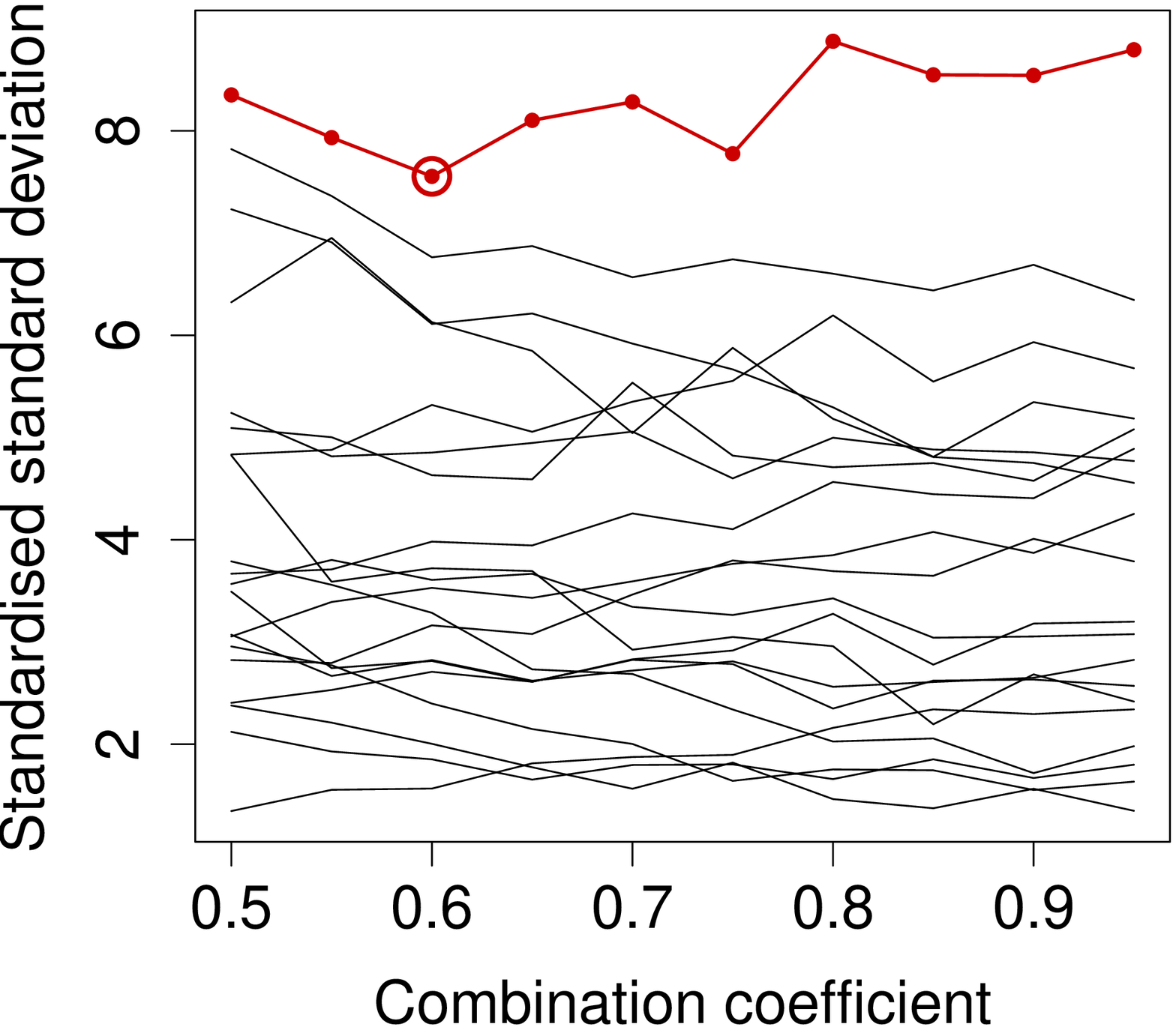}
				\subfiglabel{fig_simu_SSD_R.R3}
			}
			\caption{Results of the simulations in Series R: curves of $\critSSD$ against $\estcoefdenstinf$.}
			\label{fig_simu_SSD_R}
		\end{figure}

			In order to verify this conclusion,
			another series of simulations, Series R, is conducted.
			In this series, the true density of the infections
			is more complex than that in Series E.
			For each $i \in \nns{4}$,
			the density of infections in the $i$-th sub-square is
			$\denstinfpos{i} = \coefdenstinfpos{i} \denstpoppos{i} + (1-\coefdenstinfpos{i}) \denstdiagpos{i}$,
			where $\coefdenstinfpos{i} \in [0,1]$,
			and $\denstpoppos{i}$ and $\denstdiagpos{i}$ are
			the densities of population and cases in the $i$-th sub-square,
			respectively.
			Series R also contains three groups of simulations, R1 to R3.
			The settings are shown in \reftab{tab_simu_coef_setting_R},
			in which there are $20$ independent settings
			for $\coefdenstinfpos{1}, \omd, \coefdenstinfpos{4}$ in each group,
			and the corresponding results are given in \reffig{fig_simu_SSD_R}.
			For example, in Group R1,
			each setting of $\denstinf$ is obtained by
			generating $\coefdenstinfpos{1}, \omd, \coefdenstinfpos{4} \stackrel{\IID}{\sim} \DistrU[0.02, 0.98]$.
			Under each setting of $\denstinf$,
			$\estcoefdenstinf$ takes the $19$ values in $\{0.05, 0.1, 0.15, \omd, 0.9, 0.95\}$ one by one,
			and the corresponding values of $\critSSD(\estcoefdenstinf)$ are calculated,
			and they form a black curve in \reffig{fig_simu_SSD_R.R1}.
			The robust setting of $\estcoefdenstinf$ is the one that minim\UKUS{is}es the maximum $\critSSD$,
			i.e., the one that minim\UKUS{is}es the red bold curve in \reffig{fig_simu_SSD_R.R1},
			which is marked out by a red circle.
			The three subfigures in \reffig{fig_simu_SSD_R}
			seem more tanglesome than those in \reffig{fig_simu_SSD_E},
			which is caused by the complex setting of $\denstinf$ in Series R.
			However,
			the phenomenon presented in \reffig{fig_simu_SSD_R}
			is similar to that in \reffig{fig_simu_SSD_E}, i.e.,
			in each subfigure,
			the red bold curve becomes high
			when $\estcoefdenstinf$ is near the end of the domain of the $\coefdenstinfpos{i}$'s,
			but becomes low
			when $\estcoefdenstinf$ is around the mid-point of that domain.
			It indicates again that
			the mid-point of the possible range of $\coefdenstinfpos{i}$'s
			is a robust setting of $\estcoefdenstinf$.
			Some additional simulations in the Supplementary Material 
			also present the similar phenomenon clearly.
			Therefore, according to the above simulations under different settings of $\coefdenstinf$,
			when a possible range of the value of $\coefdenstinf$ is available,
			we should set $\estcoefdenstinf$ around the mid-point of that range,
			which is robust in the sense that
			the maximum value of $\critSSD$ over all possible settings of $\denstinf$ would not be very large.
			

	\begin{table}[!t]
			\centering
			\caption{Settings of $\denstinf$ and $\estcoefdenstinf$ in the simulations in Series R.}
			\begin{tabular*}{\textwidth}{@{\extracolsep\fill}cccc@{\extracolsep\fill}}
				\toprule
				& Group R1 & Group R2 & Group R3 \\
				\midrule
				$\coefdenstinfpos{i}$'s & $[0.02, 0.98]$ & $[0.02, 0.53]$ & $[0.47, 0.98]$ \\
				$\estcoefdenstinf$ & $\{0.05, 0.1, \omd, 0.9, 0.95\}$
				& $\{0.05, 0.1, \omd, 0.45, 0.5\}$ & $\{0.5, 0.55, \omd, 0.9, 0.95\}$ \\
				\bottomrule
			\end{tabular*}
			\label{tab_simu_coef_setting_R}
		\end{table}

		\subsection{Comparisons}\label{sect_simu_comp}
			
			In this subsection,
			we show some comparisons for the GLS against other popular methods,
			including the SIR, MH and the stratified sampling.
			In this simulation,
			the region $\region = [0,1]^{2}$ is divided into $16$ equal-sized sub-squares
			(which is similar to the previous subsection)
			within each of which
			the density of the population is uniform,
			while the densities of the cases and infections are proportional to
			a normal distribution.
			In each of the $16$ sub-squares,
			the population is fixed,
			but the numbers of cases and infections are generated randomly.
			Please refer to the Supplementary Material 
for more details about the settings. 			As for the sampling methods,
			we use the same sampling strategy in Algorithm \ref{algo_sample},
			except for the sampler in Step \ref{algo_sample_position}.
			In our proposed method,
			the size of the uniform design used in the GLS $\sizeUD = 210$.
			The proposal distributions in the SIR and MH samplers are set as
			the uniform distribution on $\region = [0,1]^{2}$,
			and the normal distribution with covariance matrix equal to
			$\infrac{1}{16} \cdot \MatI_{2}$,
			respectively.
			In the three methods GLS, SIR and MH,
			the number of sampling positions $\sizepos = 16$,
			and the value of $\estcoefdenstinf$ is set $0$ so that $\denstsam \propto \denstdiag$.
			In the stratified sampling,
			the strata are the $16$ sub-squares
			and the Neyman allocation is used to determine the sample sizes in each stratum.
			The total population $\totalpop = 8000 \times 10^{4}$
			and the total sample size $\sizesam = 1 \times 10^{4}$.
			
\begin{figure}[!t]
			\centering
			\subfigure[Relative bias]{
				\includegraphics[width=0.3\textwidth]{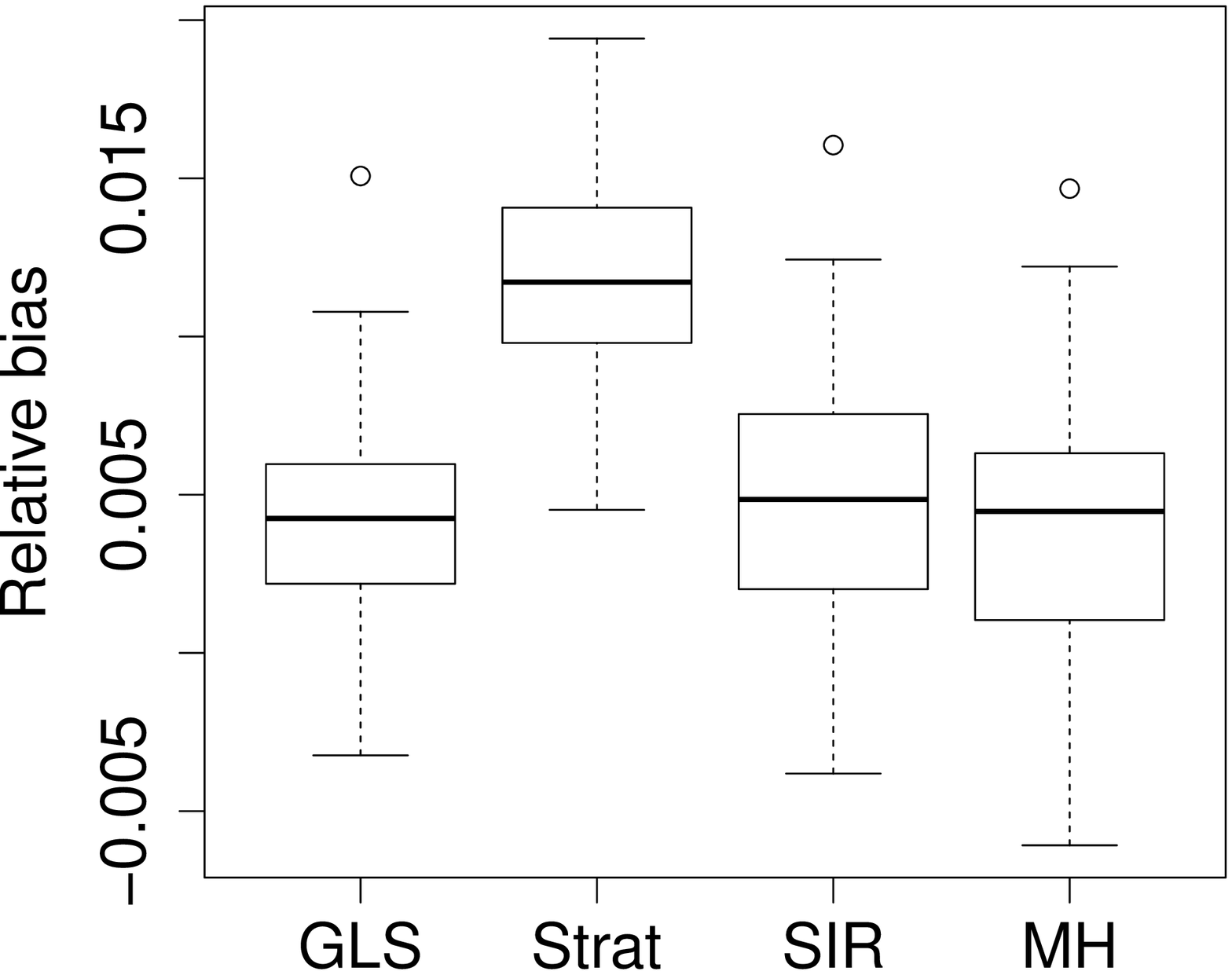}
				\subfiglabel{fig_simu_comp.bias}
			}
			\hspace{-0.05\textwidth}
			\subfigure[Standard\UKUS{is}ed standard deviation]{
				\includegraphics[width=0.3\textwidth]{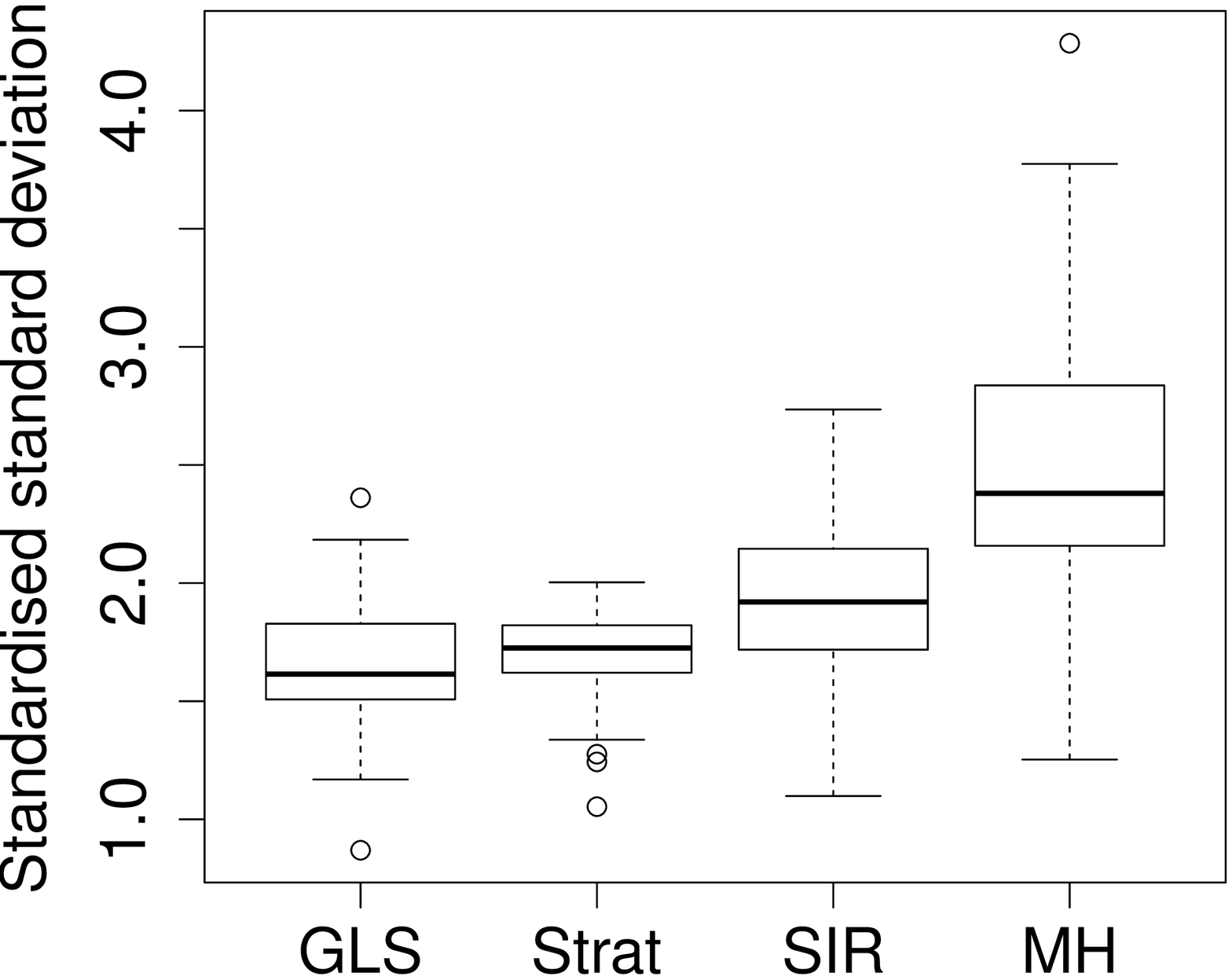}
				\subfiglabel{fig_simu_comp.SSD}
			}
			\hspace{-0.05\textwidth}
			\subfigure[Coverage rate of CI]{
				\includegraphics[width=0.3\textwidth]{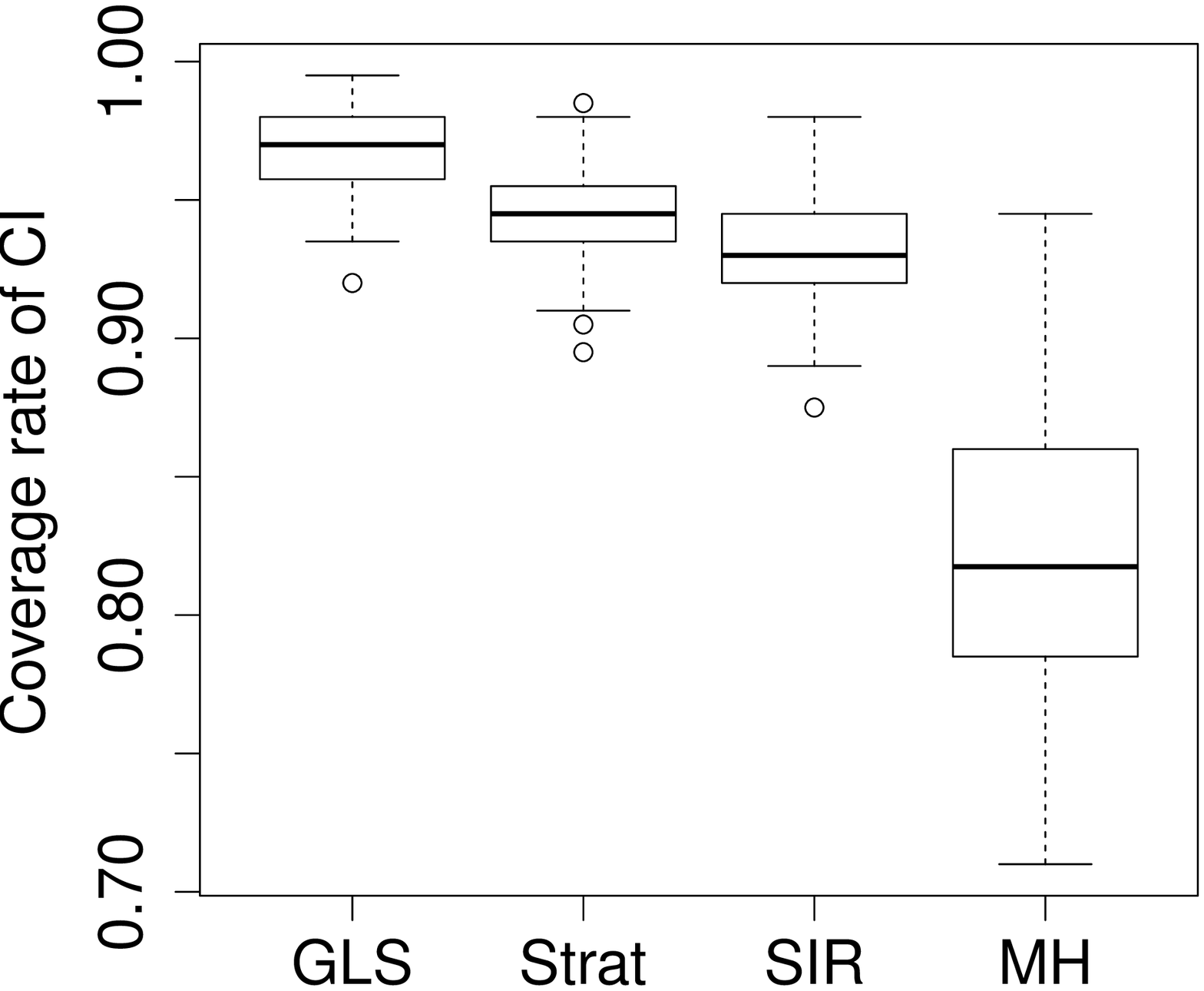}
				\subfiglabel{fig_simu_comp.CI}
			}
			\caption{Comparisons of different sampling methods.}
			\label{fig_simu_comp}
		\end{figure}

			We use the following three criteria to compare their performances:
			\begin{enumerate}
				\item Relative bias:
					the ratio of the bias of $\esttotalinf$ to the true value of $\totalinf$;
				\item Standard\UKUS{is}ed standard deviation:
					the ratio of
					the standard deviation of the estimator of $\totalinf$,
					obtained by a particular method,
					to that obtained by our method with the exact optimal settings,
					which is similar to that in \refsect{sect_simu_coef};
				\item Coverage rate of CI:
					the coverage rate of the approximate $95\%$ CI.
			\end{enumerate}
			We use $100$ different random settings of $\denstinf$ and $\denstdiag$.
			Under each of the $100$ settings, for each method,
			the values of these three criteria are calculated
			through $200$ independent simulations.
			The results are presented in \reffig{fig_simu_comp}.
			The unbiasedness of the estimators of $\totalinf$ obtained by these four methods
			is verified through \reffig{fig_simu_comp.bias}.
			From \reffig{fig_simu_comp.SSD},
			we can find that our method has the smallest $\critSSD$,
			i.e., the minimum variance of the estimator of $\totalinf$.
			The main reasons are
			that the performance of the SIR sampler depends heavily on
			the quality of the initial samples from the proposal distribution,
			that the MH sampler is easy to get stuck at some peak of the density function,
			and that the stratified sampling method
			does not util\UKUS{is}e the information about the population and the cases sufficiently.
			From \reffig{fig_simu_comp.CI},
			we can also find that our method has
			the highest coverage rate of the approximate $95\%$ CI of $\totalinf$.
			Therefore, we can conclude that our method is efficient,
			in the sense that
			its estimator of $\totalinf$ is unbiased,
			with a small variance and a high coverage rate of the CI.
			
	
	\section{A Practical Example}\label{sect_USA}
	
\begin{figure}[!t]
			\centering
			\subfigure[Density of population]{
				\includegraphics[width=0.45\textwidth]{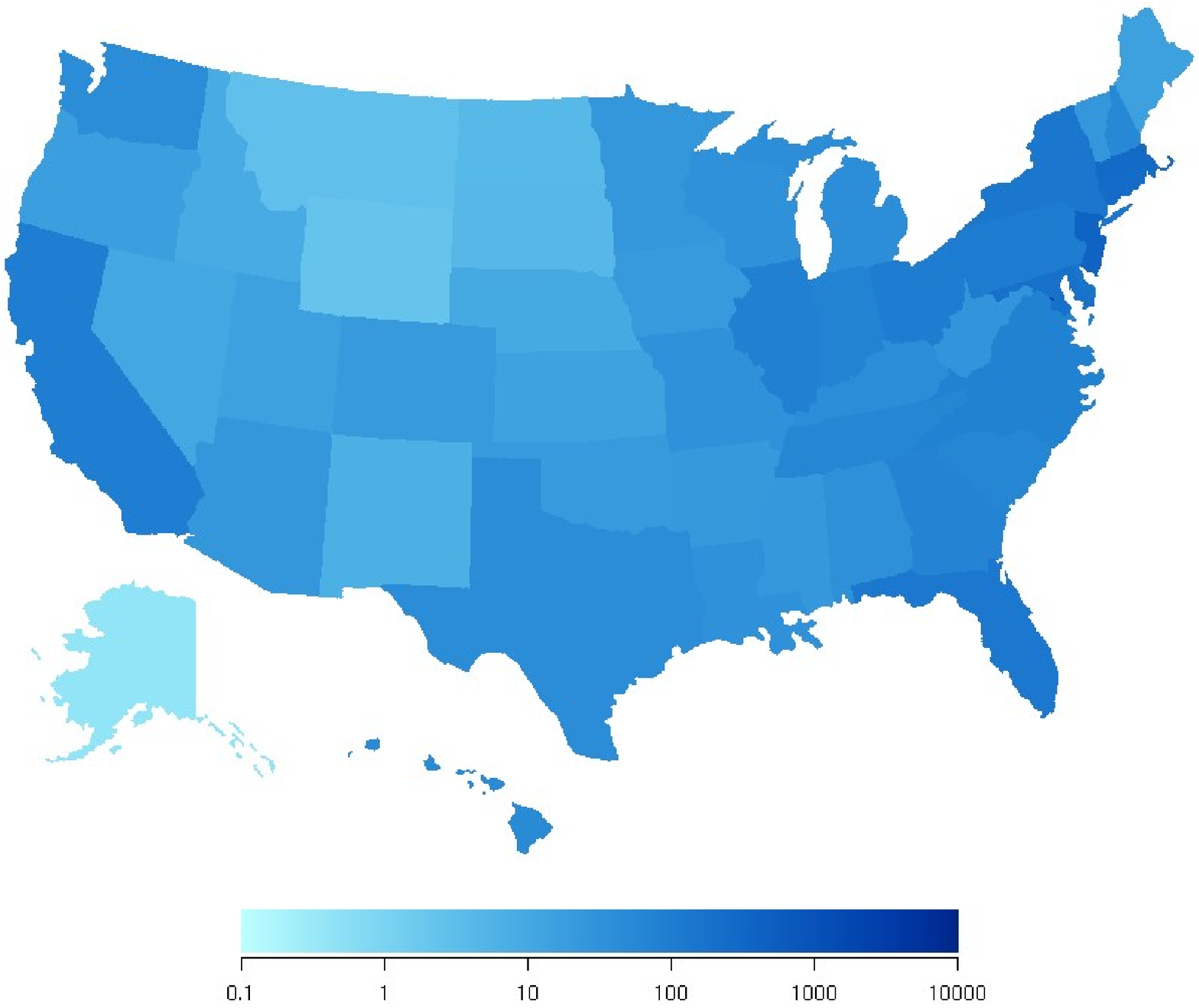}
				\subfiglabel{fig_USA_state.pop}
			}
			\subfigure[Density of cases]{
				\includegraphics[width=0.45\textwidth]{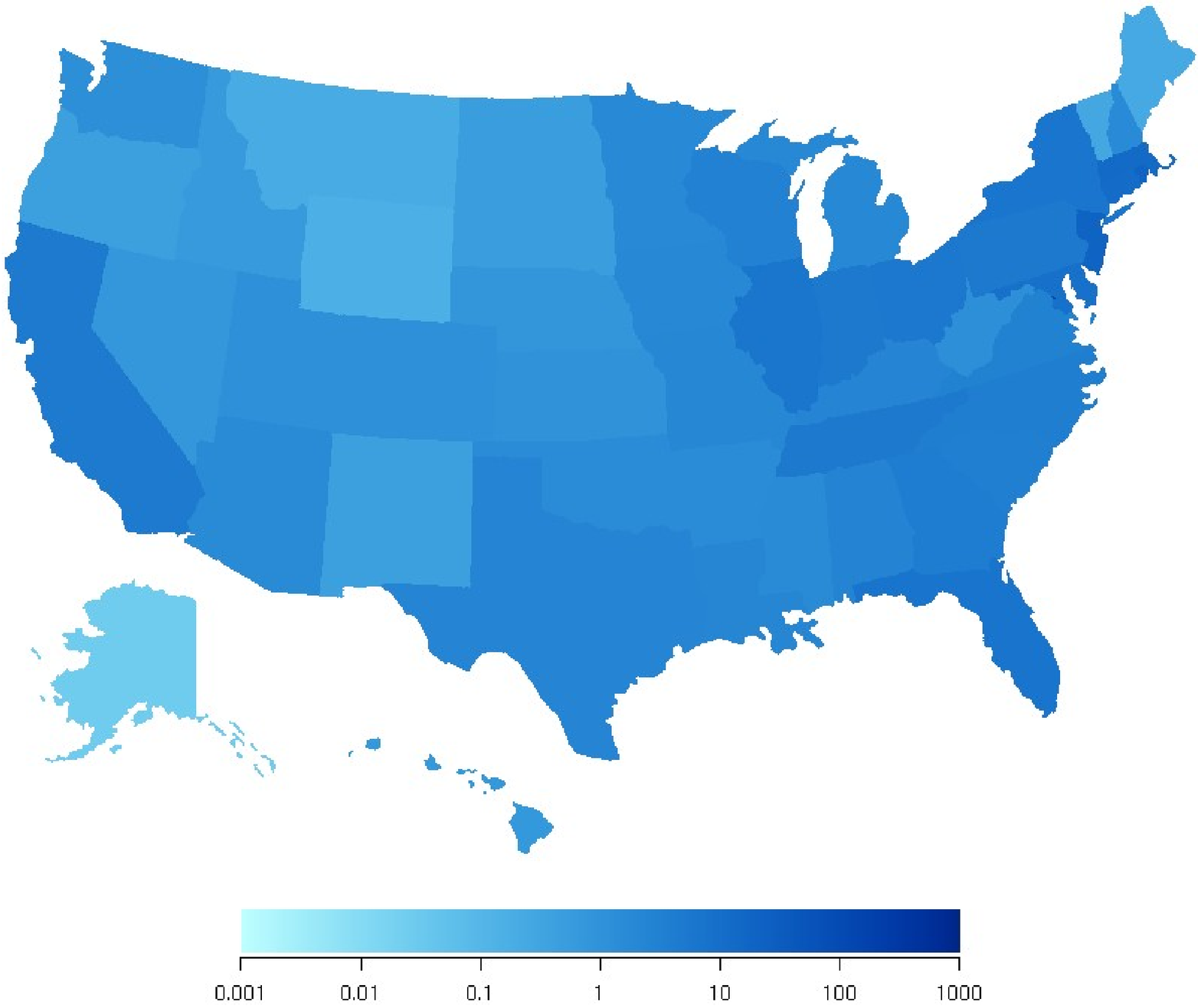}
				\subfiglabel{fig_USA_state.diag}
			}
			\subfigure[Density of infections]{				
\includegraphics[width=0.45\textwidth]{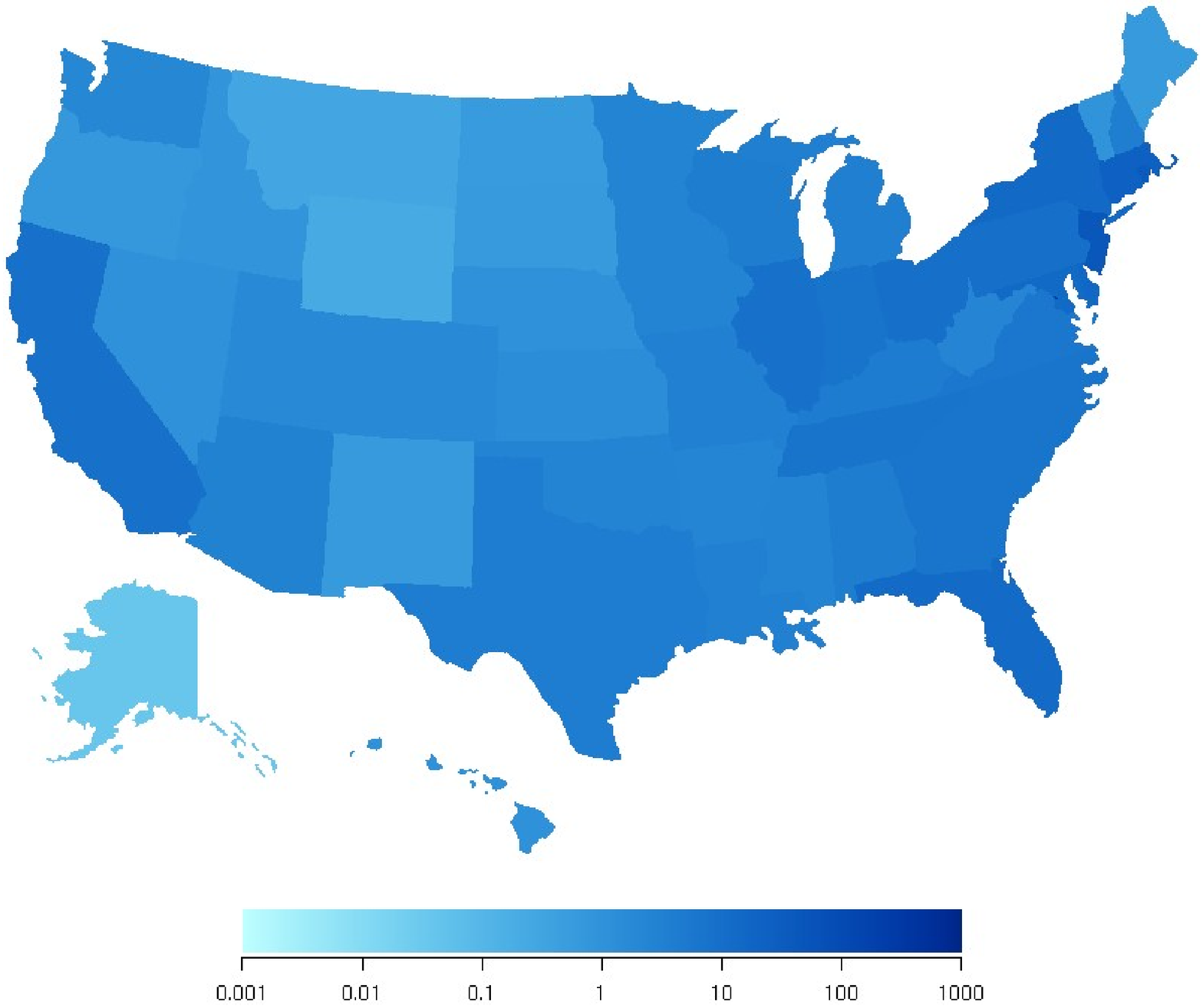}
				\subfiglabel{fig_USA_state.inf}
			}
			\subfigure[Result of one round of simulation]{
				\includegraphics[width=0.45\textwidth]{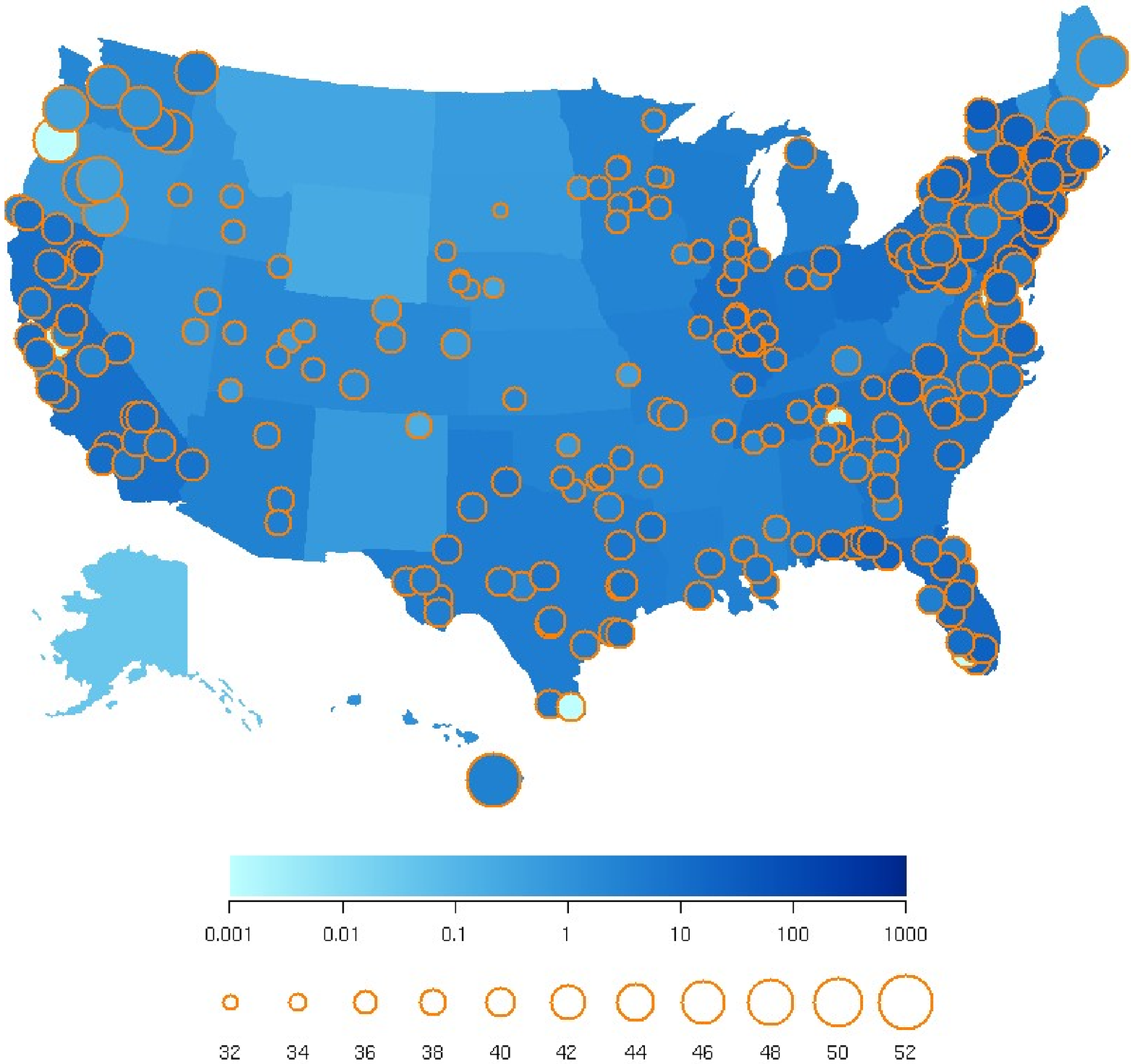}
				\subfiglabel{fig_USA_state.sample}
			}
			\caption{\ref{sub@fig_USA_state.pop}--\ref{sub@fig_USA_state.inf}:
				The densities of population, cases and infections;
				\ref{sub@fig_USA_state.sample}: the result of one round of simulation.}
			\label{fig_USA_state}
		\end{figure}

		To further illustrate the two-stage sampling strategy,
		an example based on the situation of COVID-19 in the USA is presented in this section.
		The `USA' mentioned in this section refers to
		the 50 states of the USA, as well as the District of Columbia,
		excluding other territories of the USA.
		The cumulative numbers of diagnosed cases of the 51 administrative districts in the USA
		can be obtained from \citeprefix{the}{}{the \citeauthor} \cite{2020_CDC_web}.
		Assume that
		the true densities of population $\denstpop$, cases $\denstdiag$ and infections $\denstinf$
		are the densities of
		population, cases up to \USAdatediag{}, and cases up to \USAdateinf{} in the USA, respectively.
		The corresponding densities are depicted in
		\reffig{fig_USA_state}\ref{sub@fig_USA_state.pop}--\ref{sub@fig_USA_state.inf},
		with the unit being $\mathrm{km}^{-2}$,
		and the specific data are shown in the Supplementary Material. 
		The total population and the numbers of cases and infections are
		$\totalpop = 331.319 \times 10^{6}$,
		$\totaldiag = 29.701 \times 10^{6}$ and $\totalinf = 31.467 \times 10^{6}$, respectively.
		The total sample size $\sizesam$ is set as $10000$,
		which is quite small compared to the total population of the USA.
		For comparison,
		we use both our method and the classical stratified sampling
		to estimate the total number of infections $\totalinf$.
		Based on the settings of $\totalpop$ and $\totaldiag$,
		it is reasonable to assume that the values of the true $\coefdenstinf$ are in the range $[0,0.1]$,
		and we set $\estcoefdenstinf = 0.05$ to obtain the initial rough estimation of $\denstinf$.
		In our method,
		we choose $\sizepos=250$ and $\sizeUD = 210$, as recommended in the previous sections,
		and the settings of the sampling density $\denstsam$ and the allocation method of the sample sizes
		are nearly optimal.
		As for the stratified sampling,
		the strata are defined as the 51 administrative districts,
		which is consistent with the area partition when recoding the epidemic-related data;
		the Neyman allocation \citep{2019_Lohr_BrooksCole} was used.
		The details about the stratified sampling are described in
the Supplementary Material. 
		

		The results of $200$ rounds of independent simulations are shown in \reftab{tab_USA_comp}.
		Though the standard deviation of the estimator obtained by our proposed method
		is slightly smaller than that of the stratified sampling,
		our method has higher coverage rate of CI.
		Note that in practice,
		the samples must be drawn from several selected sampling positions in each stratum.
		Due to this,
		the stratified multi-stage sampling is often used \citep{bib_pp_36},
		which can further increase the variance of the estimator.
		For the stratified multi-stage sampling method,
		big efforts have to be made in order to adapt to a complicated distribution of population,
		and the variance may become difficult to estimate.
		However, our method can automatically util\UKUS{is}e the complicated information about
		the distributions of population and cases,
		while also maintaining the simplicity in estimating $\totalinf$ and its variance.
		
	
	\begin{table}[!t]
			\centering
			\caption{Results of the comparison between our method and the stratified sampling.}
			\begin{tabular*}{\textwidth}{@{\extracolsep\fill}crrrr@{\extracolsep\fill}}
				\toprule
				Method
				& \shortstack{Sample \\ mean of $\esttotalinf$}
				& \shortstack{Sample standard \\ deviation of $\esttotalinf$}
				& \shortstack{Coverage \\ rate of CI} \\
				\midrule
				Our method & $31.612 \times 10^{6}$ & $0.995 \times 10^{6}$ 
				& $99.5\%$ \\
				Stratified sampling & $31.691 \times 10^{6}$ & $1.000 \times 10^{6}$ 
				& $95.0\%$ \\
				\bottomrule
			\end{tabular*}
			\label{tab_USA_comp}
		\end{table}

		Further, we show the result of one of the $200$ rounds of simulations to explain our method.
		In this round of simulation,
		the estimated number of infections $\esttotalinf = 29.70 \times 10^{6}$,
		with the estimated standard deviation $\sqrt{\hat{v}(\esttotalinf)}=1.36 \times 10^{6}$.
		The approximate $95\%$ CI does cover the true $\totalinf$.
		The sampling positions are presented in \reffig{fig_USA_state.sample},
		where the background is the true $\denstinf$,
		and the circles represent the sampling positions,
		with their col\UKUS{our}s showing the estimated values of $\denstinf$ in $\mathrm{km}^{-2}$,
		and their diameters showing the sample sizes there.
		The final total sample size in this round is $9999$,
		smaller by $1$ than the expected $\sizesam=10000$,
		due to the round-off error at each sampling position.
		The sample sizes at these sampling positions are nearly the same,
		which is determined by the characters of $\denstpop$ and $\denstdiag$,
		as well as $\estcoefdenstinf$.
				
	\section{Conclusions}\label{sect_conclusions}
	
		In this paper,
		we propose a novel, two-stage sampling strategy
		to estimate the number of infections of a pandemic.
		Our method can sufficiently util\UKUS{is}e the information about the distributions of
		both the population and the diagnosed cases,
		which can gather information more flexibly than the existing methods,
		and hence, more efficient.
		Moreover,
		our two-stage sampling strategy does not involve any discrete structures like strata or clusters,
		therefore, it can easily and automatically adapt to
		the complicated distributions of population and cases,
		and the corresponding estimating method keeps simple.
		The GLS algorithm used in our method is also easy to implement,
		since it does not need a proposal distribution.
		Its performance is robust against
		the complexity and multimodality of the sampling density,
		which overcomes the drawbacks of the other popular samplers for general probability densities,
		such as the SIR or MH algorithm.
		
		Since the true density of the infections is not known in practice,
		obtaining the exact optimal settings of
		the sampling density and the allocation function in this sampling strategy is unrealistic.
		Instead,
		we discuss the nearly optimal ones for practical implementation,
		which is based on an initial rough estimate of the density of infections.
		The total sample size can be determined by the pre-defined estimation precision.
		The small sample sizes at the sampling positions can be avoided
		by modifying the allocation of sample sizes based on a convex combination.
		In addition,
		in the second stage of our method,
		we just consider the simple random sampling method for simplicity.
		To further improve the efficiency and eliminate the selection bias,
		other sampling methods like the stratified sampling can be taken into account,
		with some slight modifications for the corresponding formulae in the second stage.
		Further, by the numerical simulations,
		we discuss the robust setting of the combination coefficient
		for the rough estimator of the density of infections in the minimax sense.
		We also compare the GLS algorithm with the SIR, MH and stratified sampling methods
		in terms of the relative bias, standard deviation and coverage rate of CI.
		It shows that the GLS algorithm
		has the smallest standard deviation and the highest coverage rate of the approximate 95\% CI.
		Moreover, we apply our method to the investigation of COVID-19 in the USA.
		Our method shows good performance and has higher coverage rate of CI
		compared with the stratified sampling.
		Hence,
		these simulations and the practical example
		verify the efficiency of our proposed two-stage sampling method,
		whatever the sampling density is.

	\section*{Acknowledgments}
		
		This work was supported by
		the National Natural Science Foundation of China under Grant Nos. 11871288, 11771220 and 12131001;
		and Natural Science Foundation of Tianjin under 19JCZDJC31100.
		The first two authors contributed equally to this work.
		
		


\end{document}